\documentclass[lengthcheck,superscriptaddress]{revtex4-2}
\usepackage[T1]{fontenc}
\usepackage[utf8]{inputenc}
\usepackage{color}
\usepackage{babel}
\usepackage{array}
\usepackage{multirow}
\usepackage{amsmath}
\usepackage{amssymb}
\usepackage{graphicx}
\usepackage[unicode=true,pdfusetitle,
 bookmarks=true,bookmarksnumbered=false,bookmarksopen=false,
 breaklinks=false,pdfborder={0 0 1},backref=false,colorlinks=true]
 {hyperref}

\makeatletter

\providecommand{\tabularnewline}{\\}

\newenvironment{lyxlist}[1]
	{\begin{list}{}
		{\settowidth{\labelwidth}{#1}
		 \setlength{\leftmargin}{\labelwidth}
		 \addtolength{\leftmargin}{\labelsep}
		 }}
	{\end{list}}

\makeatletter
\renewcommand{\p@subsection}{}
\renewcommand{\p@subsubsection}{}
\makeatother

\makeatother

\begin{document}
\title{Cumulant expansion framework for internal gradient distributions tensors}
\author{Leonardo A. Pedraza Pérez}
\affiliation{Instituto Balseiro, CNEA, Universidad Nacional de Cuyo, S. C. de Bariloche,
8400, Argentina.}
\affiliation{Centro Atómico Bariloche, CNEA, S. C. de Bariloche, 8400, Argentina.}
\author{Gonzalo A. Álvarez}
\altaffiliation{gonzalo.alvarez@cab.cnea.gov.ar}

\affiliation{Instituto Balseiro, CNEA, Universidad Nacional de Cuyo, S. C. de Bariloche,
8400, Argentina.}
\affiliation{Centro Atómico Bariloche, CONICET, CNEA, S. C. de Bariloche, 8400,
Argentina.}
\affiliation{Instituto de Nanociencia y Nanotecnologia, CNEA, CONICET, S. C. de
Bariloche, 8400, Argentina}
\begin{abstract}
Magnetic resonance imaging is a powerful, non invasive tool for medical
diagnosis. The low sensitivity for detecting the nuclear spin signals,
typically limits the image resolution to several tens of micrometers
in preclinical systems and millimeters in clinical scanners. Other
sources of information, derived from diffusion processes of intrinsic
molecules as water in the tissues, allow getting morphological information
at micrometric and submicrometric scales as potential biomarkers of
several pathologies. Here we consider extracting this morphological
information by probing the distribution of internal magnetic field
gradients induced by the heterogeneous magnetic susceptibility of
the medium. We use a cumulant expansion to derive the dephasing on
the spin signal induced by the molecules that explore these internal
gradients while diffuse. Based on the cumulant expansion, we define
internal gradient distributions tensors (IGDT) and propose modulating
gradient spin echo sequences to probe them. These IGDT contain microstructural
morphological information that characterize porous media and biological
tissues. We evaluate the IGDT effects on the magnetization decay
with typical conditions of brain tissue and show their effects can
be experimentally observed. Our results thus provide a framework for
exploiting IGDT as quantitative diagnostic tools.
\end{abstract}
\maketitle

\section{Introduction}

Magnetic resonance imaging (MRI) has proven to be a powerful technique
for non-invasive medical diagnosis. The implementations of MRI techniques
usually limit the spatial resolution to millimeters in clinical scanners
and to tens of micrometers in preclinical systems. However, the monitored
nuclear spins can probe smaller spatial scales. Different methods
thus exploit various physical processes sensed by the nuclear spins
such as molecular diffusion \citep{Callaghan1995,Latour_1995,Stepisnik2006,Assaf2008,Gore2010,Ong_2010,Komlosh_2011,Shemesh2013,Xu2014,Shemesh2015,Drobnjak_2015,Palombo_2016,Nilsson_2017,Drake2018,Novikov2018,Novikov_2018,Xu_2019,Huang2020,Veraart_2020,Palombo_2020,Capiglioni2021}
or tissue properties as the magnetic susceptibility \citep{Huerlimann1998,Song2000a,Han2011,Liu2011,Duyn2013,Alvarez2014,Zhang2016,Xu_2017,Alvarez2017,Sandgaard2022}
to generate contrast that can unveil the tissue microstructure.

Based on these principles, diffusion-weighted images (DWI) display
information about the motion of water molecules within tissues \citep{Callaghan2011}.
DWI has become a common diagnostic tool in clinical practice. For
instance, the most established clinical indication for DWI is the
assessment of cerebral ischemia \citep{Drake2018}. DWI has been also
used to monitor lesion aggressiveness and tumor response in oncology
imaging \citep{Messina_2020}. Its extension to diffusion tensor imaging
(DTI), based on estimating the components of the apparent diffusion
coefficient along multiple directions, enables to determine tissue
microstructure anisotropies which are very useful for tracking brain
connectivity \citep{Mori1999,Basser1994,Basser1994a,Basser2002,LeBihan2001,LeBihan2003}.
Nevertheless, there are cases where the anisotropy of the apparent
diffusion tensor is small, as occurs in wide compartments or in fiber
crossing regions in brain, and the application of DTI it is thus limited
\citep{Basser2002,Mori2002,Jones_1999,Frank_2001}. 

Alternatively, susceptibility-weighted imaging (SWI) exploits the
effect of tissues magnetic susceptibility variations to generate high
resolution (0.3 mm) anatomical images in high field scanners (>7 Tesla)
\citep{Duyn2013}. Tissue susceptibility variations are encoded in
the amplitude and phase of the magnetic resonance (MR) signal. Using
this information it is possible to suppress or enhance spectral components
and modify the contrast of different tissues. These concepts has been
used for water/fat separation \citep{Haacke1986,Mao_1993}, MR angiography
\citep{Axel1986,Alfidi_1987,Reichenbach1997} and gray/white-matter
contrast \citep{Duyn_2007}. Its application as disease biomarkers
is highly encouraging for observing image changes based on iron and
calcium concentration variations, and for axon demyelination \citep{Eskreis_Winkler_2014,Eskreis_Winkler_2016,Wang_2022}.
However, to attain this quantitative information it is often required
the rotation of the studied subject \citep{Liu2009}.

Biological tissues have very complex structures and compartmentalization
heterogeneities due to different molecular compositions. Its intrinsic
heterogeneity is imitated by magnetic susceptibility changes along
the tissue structure \citep{Haacke_2004,Liu_2010,Alvarez2014,Alvarez2017,Sandgaard2022}.
In presence of an external magnetic field, the susceptibility discontinuities
induce internal magnetic field gradients \citep{Huerlimann1998,Nestle_2002,Han2011,Liu2011,Lee2012,Liu_2014}.
These internal or background gradients are correlated with the pore
size distribution, thus allowing the inference of tissue microstructure
sizes from the internal gradients statistics \citep{Song2000,Song2000a,Chen_2003,Kuntz_2007,Cho_2012,Zhang2016}.
Moreover, the internal gradient distributions contain also information
about microstructure anisotropy, for example, related with fiber orientations
\citep{Han2011,Alvarez2017}. These techniques are specially useful
when DTI fails due to almost isotropic diffusion scenarios. Importantly,
compared with quantitative SWI, the statistical characterization of
internal gradient distribution can be done without reorienting the
subject with respect to the direction of the main magnetic field,
thus allowing to extract anisotropic information based on susceptibility
heterogeneities within tissues \citep{Han2011,Alvarez2017}.

Brain physiology is also regulated by an assemblage of molecules and
structures as the myelin sheath of axons, with significant magnetic
susceptibility changes with respect to the surrounding medium. The
degree of axon myelination has been shown to significantly affect
the transverse relaxation time and spin phases in white matter \citep{Liu2011,Lee2012}.
Therefore, it is expected that internal gradient distributions also
show correlation with the amount of myelin in such tissues \citep{Duyn2013,Chen_2013,Xu_2017,Fajardo},
thus being a potential biomarker of degenerative diseases.

In this article, we provide a perturbative framework to characterize
and determine internal (or background) gradient distributions by its
first statistical moments. These moments define internal gradient
distributions tensors (IGDT) that provide information about the media
anisotropy. We formalize previous works that evidence these IGDT \citep{Cho_2009,Han2011,Cho_2012,Alvarez2017,Fajardo},
based on a cumulant expansion of the spin dephasing beyond a Gaussian
phase approximation. We consider an internal gradient ensemble model,
where spin-bearing molecules diffuse in presence of an effective gradient.
This effective gradient is determined by the average gradient seen
by the spin along the diffusion trajectory. This approach allows simplifying
high order correlations of the spin displacement as the molecular
diffusion can be considered a Gaussian process, and then write the
magnetization signal with an IGDT expansion of different ranks. The
IGDT may have subtle effects on the magnetization signal decay, thus
we exploit the IGDT expansion to design Modulated Gradient Spin-Echo
(MGSE) sequences \citep{Callaghan1995,Callaghan2011} that use cross-correlations
between an applied and the internal gradient to enhance their contributions
to the MR signal \citep{Alvarez2017}. These MGSE sequences enconde
the internal gradient information on the diffusion weighting decay
rather than on a phase-shift in the magnetization signal, as the latter
is removed. We perform a combination of analytical and numerical
analyses to demonstrate the feasibility of determining the IGDT in
realistic conditions using typical brain tissue properties. We also
discuss the validity of the proposed framework.

In Sec. \ref{sec:Spins-diffusing-in} we describe the physical problem
and the considered internal gradient ensemble model. Section \ref{sec:Cumulant-expansion-framework}
introduces the cumulant expansion framework for the magnetization
decay and the derivation of the IGDT expansion. In Sec. \ref{sec:Improving-IGDT-eff}
we propose a method for measuring the IGDT and design MGSE sequences
to enhance the IGDT effects on the magnetization signal decay. Section
\ref{sec:IGDT-in-brain} discusses the feasibility of measuring the
IGDT for conditions similar to those founds in brain tissue. In Sec.
\ref{sec:scop-limitations} we discuss the scope and limitations of
the considered model for characterizing the internal gradients. Finally,
in Sec. \ref{sec:Conclusions} we summarize the results and elaborate
some conclusions and outlook.

\section{Spins diffusing in an inhomogeneous field\label{sec:Spins-diffusing-in}}

\subsection{Spin phase modulated by MGSE sequences and internal gradients}

We consider an heterogeneous medium in presence of a static magnetic
field $\boldsymbol{B}_{0}$ in the $z$ direction. Due to inhomogenities
of the magnetic susceptibility in the sample, the effective magnetic
field becomes also inhomogeneous leading to local field variations
$\Delta\boldsymbol{B}_{0}(\boldsymbol{x})$ depending on the spatial
position $\boldsymbol{x}$. The typical magnetic field variations
induced by the magnetic susceptibility changes in porous media are
on the order of $\approx10^{-6}\boldsymbol{B}_{0}$ \citep{Wharton2010,Han2011,Callaghan2011,Connolly_2019}.
As these local field variations are significantly weaker than the
static magnetic field strength, we only need to consider the $z$
component of that local field, $\Delta B_{0}(\boldsymbol{x})$. Since
$\Delta B_{0}(\boldsymbol{x})$ intrinsically depends on the microstructure
morphology of the medium, we are interested to probe it to extract
sub-voxel morphological information. As these field variations originate
internal or background gradients $\boldsymbol{G}_{0}(\boldsymbol{x})=\nabla\Delta B_{0}(\boldsymbol{x})$,
with $\nabla$ being the differential operator with respect to $\boldsymbol{x}$,
they can be probed with the nuclear spin-bearing molecules diffusing
within these internal field gradients \citep{Rochefort2010,Sen1999,Wharton2010,Li2012,Alvarez2014,Alvarez2017,Sandgaard2022}.

The Brownian motion of spin-bearing particles in a constant magnetic
field gradient induces spin dephasing that cannot be fully refocused
by spin-echo based sequences \citep{Hahn1950,Carr1954,Meiboom1958}.
This dephasing thus leads to a magnetization decay that is typically
used to probe the molecular diffusion in porous media \citep{Stejskal1965,LeBihan2003,Grebenkov2007,Jones2010,Callaghan2011}.
The temporal diffusion process can be characterized by applying magnetic
field gradients that vary over time with a modulation function $f_{G}(t)$
\citep{Stepisnik1993,Stepisnik2006,Lasic2006,Gore2010,Alvarez2013,Shemesh2013}.
Thus, the phase acquired by a diffusing spin-bearing particle during
a diffusion time $t_{d}$ depends on the spin-bearing particle trajectory
$\boldsymbol{x}(t)$, the internal field $\Delta B_{0}$ and the applied
gradient $\boldsymbol{G}$ \citep{Sen1999,Lasic2006,Grebenkov2007,Han2011,Li2012,Rochefort2010,Alvarez2014,Alvarez2017}
as 
\begin{align}
\phi\left[t_{d},\boldsymbol{x}(t),\Delta B_{0}\right]=\gamma\int_{0}^{t_{d}}dt' & \left\{ \boldsymbol{x}(t')\cdot\boldsymbol{G}f_{G}(t')\right.\nonumber \\
 & \left.+f_{0}(t')\Delta B_{0}[\boldsymbol{x}(t')]\right\} ,\label{eq:exact_phase}
\end{align}
where $\gamma$ is the gyromagnetic ratio of the nucleus.

\begin{figure}
\includegraphics[width=0.9\columnwidth]{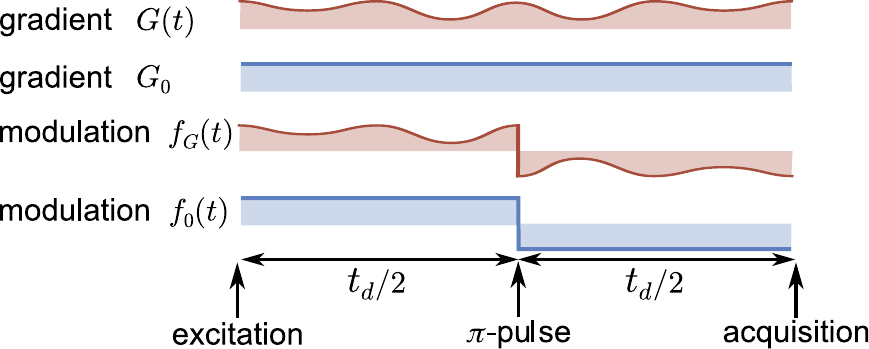}

\caption{\label{fig:simple_modulations} Scheme for the effective applied gradient
modulation function $f_{G}(t)$ and internal gradient modulation $f_{0}(t)$,
based on an arbitrary applied gradient waveform $G(t)$ and a Hahn-echo
sequence for modulating the internal gradient $G_{0}$. The RF $\pi$-pulse
of the Hahn-echo sequence induces a sign inversion of the applied
and internal gradient modulations.}
\end{figure}

The sign of the internal field can be controlled by radio-frequency
(RF) $\pi$-pulses and it is described by the modulation function
$f_{0}(t)$. The RF $\pi$-pulses globally affect the phase evolution
of the spins. Each of these pulses effectively switch also the sign
of the applied gradient in the toggling frame representation of the
spins \citep{Alvarez2014,Alvarez2017}. The considered modulation
$f_{G}(t)$ includes the effects of the total modulation due to the
applied gradient control and the effective sign switches induced by
the RF $\pi$-pulses. Thus, $f_{G}(t)$ denotes the effective modulation
of the applied gradient interaction with the spins, see Fig. (\ref{fig:simple_modulations}).
Here we consider MGSE sequences \citep{Callaghan1995,Callaghan2011},
where the modulating functions satisfy 
\begin{equation}
\int_{0}^{t_{d}}dt'f(t')=0.\label{eq:MGSE_modulation_def}
\end{equation}

\subsection{Internal Gradient ensemble model\label{subsec:Internal-Gradient-ensemble}}

\begin{figure}
\includegraphics[width=1\columnwidth]{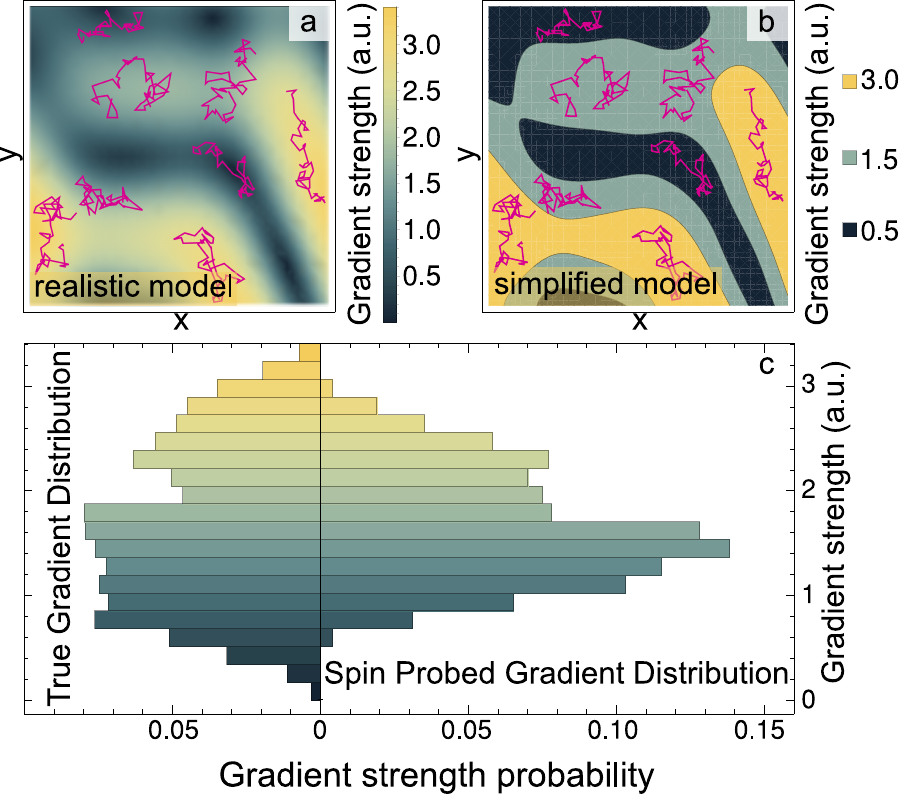}

\caption{\label{fig: model_scheme}Schematic representation of the internal
gradient ensemble model. The internal gradient strengths are shown
in colors represented in arbitrary units by the color bars. Realizations
of the stochastic path of the Brownian motion of spin-bearing particles
are shown with pink solid lines. (a) The gradient strength of a realistic
model where a spin particle can diffuse freely in an inhomogeneous
magnetic field. (b) Scheme for the simplified model where we assume
a spin-bearing particle moves in a effective and constant gradient.
(c) Comparison between the true gradient distribution associated to
panel (a) and the effective gradient distribution probed by the diffusing
spin-bearing particles schematized in (b). For the latter, we consider
uniformly distributed spins that diffuse during a time $t_{d}$ achieving
a root mean square displacement $\sqrt{\left\langle \left(\boldsymbol{x}(t)-\left\langle \boldsymbol{x}\right\rangle \right)^{2}\right\rangle }$
much lower than the boundary length of the simulation matrix. The
constant gradient distribution mimicks the true internal gradient
distribution.}
\end{figure}

An exact and rigorous treatment that describes the phase accumulation
of Eq. (\ref{eq:exact_phase}) for a spin ensemble is in general very
complex. We thus assume that the spin phase is characterized by an
effective and spatially constant internal field gradient $\boldsymbol{G}_{0}$
during the diffusion time $t_{d}$ (Fig. \ref{fig: model_scheme}).
Then, the spin phase is simplified to the form
\begin{align}
\phi\left[t_{d},\boldsymbol{x}(t),\boldsymbol{G}_{0}\right]=\gamma\int_{0}^{t_{d}}dt'\boldsymbol{x}(t')\cdot & \left\{ \boldsymbol{G}f_{G}(t')\right.\nonumber \\
 & \left.+\boldsymbol{G}_{0}f_{0}(t')\right\} .\label{eq:spin-phase}
\end{align}
The internal gradient $\boldsymbol{G}_{0}$ may be different for each
particle depending on their position and diffusion pathway. We thus
consider an ensemble of spin particles, and an ensemble of internal
gradients \citep{Alvarez2017,Fajardo} leading to what we call the
\emph{internal gradient ensemble model} (Fig. \ref{fig: model_scheme}).

The only relaxation mechanism of the spin phase that we consider here
is the diffusion process in presence of magnetic field gradients.
The total spin magnetization is given by the spin ensemble average
that involves all possible realization of the spin's accumulated phase.
Considering the introduced \emph{internal gradient ensemble model},
the spin ensemble average is separated in the average of all possible
fluctuating diffusion paths and the average over the distribution
of the internal magnetic field gradients as
\begin{equation}
M(t_{d})=\left\langle e^{i\phi\left[t_{d},\boldsymbol{x}(t),\boldsymbol{G}_{0}\right]}\right\rangle _{\phi}=\left\langle e^{i\phi\left[t_{d},\boldsymbol{x}(t),\boldsymbol{G}_{0}\right]}\right\rangle _{x,G_{0}}.\label{eq:magnetization_def_1}
\end{equation}
Here, the brackets $\left\langle \cdot\right\rangle _{\phi}$ and
$\left\langle \cdot\right\rangle _{x,G_{0}}$ denote the ensemble
average over the spin's accumulated phase, and the double average
over the particle displacement paths $\boldsymbol{x}(t)$ and the
distribution of internal gradients $\boldsymbol{G}_{0}$ respectively.
Figure \ref{fig: model_scheme} shows a schematic representation of
these assumptions to describe the internal gradient distributions.
A realistic model for single random realizations of spin-bearing particles
moving freely within a space with internal gradient distributions
is shown in Fig. \ref{fig: model_scheme}a. Along the Brownian motion
paths, the spins interact with different gradient strengths depending
on their spatial position. Our simplified model for this motion is
shown schematically in Fig. \ref{fig: model_scheme}b, where we assume
that, during the diffusion probing time $t_{d}$, a spin diffuses
within a spatial region represented by an effective and constant magnetic
field gradient. Then, the ensemble of constant gradient strengths
considered in Eq. (\ref{eq:magnetization_def_1}) accounts for the
distribution of particles in the different gradient field strengths.
Figure \ref{fig: model_scheme}c shows how a realistic internal gradient
distribution is mimicked by the one determined by the internal gradient
ensemble model. Further considerations of the regimes where this gradient
ensemble distribution can be used are discussed in Sec. \ref{sec:scop-limitations}.

\section{Cumulant expansion framework for internal gradient distributions\label{sec:Cumulant-expansion-framework}}

\subsection{Cumulant expansion for the spin dephasing}

The fluctuating spin phase in presence of a constant gradient typically
follows a Gaussian distribution \citep{Klauder1962,Stepisnik1999,Alvarez2013,Ziener_2018,Rotkopf_2021}.
However, due to the internal gradient distribution, the accumulated
spin phase defined in Eq. (\ref{eq:spin-phase}) deviates from the
Gaussian assumption. We thus determine the ensemble average of Eq.
(\ref{eq:magnetization_def_1}) by a cumulant expansion of the spin
phase \citep{Kubo_1962,Kampen2007}

\begin{align}
\ln M(t_{d})=-\frac{1}{2}\left\langle \left\langle \phi(t_{d})^{2}\right\rangle \right\rangle  & +\frac{1}{4!}\left\langle \left\langle \phi(t_{d})^{4}\right\rangle \right\rangle \nonumber \\
 & -\frac{1}{6!}\left\langle \left\langle \phi(t_{d})^{6}\right\rangle \right\rangle +\cdots.\label{eq:cum_exp_phi}
\end{align}
Here, the double brackets $\left\langle \left\langle \phi^{n}\right\rangle \right\rangle $
denote the $n$th order cumulant for the random variable $\phi$. 

Following the internal gradient ensemble model introduced in Sec.
\ref{subsec:Internal-Gradient-ensemble}, we separate the average
of the spatial position due to the diffusion process from the average
over the internal gradient distribution. The second order cumulant
is thus
\begin{align}
\left\langle \left\langle \phi(t_{d})^{2}\right\rangle \right\rangle = & \left\langle \Delta\phi\left[\boldsymbol{x}(t_{d}),\boldsymbol{G}_{0}\right]^{2}\right\rangle _{x,G_{0}}\nonumber \\
= & \beta_{ij}^{GG}(t_{d})G_{i}G_{j}+\beta_{ij}^{00}(t_{d})\left\langle \boldsymbol{G}_{0}\boldsymbol{G}_{0}\right\rangle _{ij}\nonumber \\
 & \hphantom{\beta_{ij}^{GG}(t_{d})G_{i}G_{j}+}+2\beta_{ij}^{0G}(t_{d})G_{i}\left\langle \boldsymbol{G}_{0}\right\rangle _{j},\label{eq:cum_2}
\end{align}
where $\Delta\phi=\phi-\left\langle \phi\right\rangle $ is the deviation
from the mean accumulated spin phase that we assume null $\left\langle \phi(t)\right\rangle =0$,
as the system is considered at the steady state under the application
of MGSE sequences that satisfy Eq. (\ref{eq:MGSE_modulation_def}).
We use the Einstein sum convention in Eq. (\ref{eq:cum_2}), where
the indexes $i,j=x,y,z$ represent the three spatial directions. The
averages $\left\langle \boldsymbol{G}_{0}\right\rangle _{j}$ and
$\left\langle \boldsymbol{G}_{0}\boldsymbol{G}_{0}\right\rangle _{ij}$
denote the matrix elements of the first two moments of the internal
gradient distribution. The attenuation matrices 
\begin{align}
\boldsymbol{\beta}^{GG}(t_{d})= & \gamma^{2}\int_{0}^{t_{d}}dt_{1}\int_{0}^{t_{d}}dt_{2}f_{G}(t_{1})f_{G}(t_{2})\nonumber \\
 & \hphantom{\gamma^{2}\int_{0}^{t_{d}}dt_{1}\int_{0}^{t_{d}}dt_{2}f_{G}}\times\left\langle \Delta\boldsymbol{x}(t_{1})\Delta\boldsymbol{x}(t_{2})\right\rangle \nonumber \\
= & \gamma^{2}\int_{\mathbb{-\infty}}^{\infty}\frac{d\omega}{2\pi}\left|F_{G}(\omega,t_{d})\right|^{2}\boldsymbol{S}(\omega),\label{eq:betaGG}
\end{align}
 
\begin{align}
\boldsymbol{\beta}^{00}(t_{d})= & \gamma^{2}\int_{0}^{t_{d}}dt_{1}\int_{0}^{t_{d}}dt_{2}f_{0}(t_{1})f_{0}(t_{2})\nonumber \\
 & \hphantom{\gamma^{2}\int_{0}^{t_{d}}dt_{1}\int_{0}^{t_{d}}dt_{2}f_{G}}\times\left\langle \Delta\boldsymbol{x}(t_{1})\Delta\boldsymbol{x}(t_{2})\right\rangle \nonumber \\
= & \gamma^{2}\int_{\mathbb{-\infty}}^{\infty}\frac{d\omega}{2\pi}\left|F_{0}(\omega,t_{d})\right|^{2}\boldsymbol{S}(\omega)\label{eq:beta00}
\end{align}
 and 
\begin{align}
\boldsymbol{\beta}^{0G}(t_{d})= & \gamma^{2}\int_{0}^{t_{d}}dt_{1}\int_{0}^{t_{d}}dt_{2}f_{0}(t_{1})f_{G}(t_{2})\nonumber \\
 & \hphantom{\gamma^{2}\int_{0}^{t_{d}}dt_{1}\int_{0}^{t_{d}}dt_{2}f_{G}}\times\left\langle \Delta\boldsymbol{x}(t_{1})\Delta\boldsymbol{x}(t_{2})\right\rangle \nonumber \\
= & \gamma^{2}\int_{\mathbb{-\infty}}^{\infty}\frac{d\omega}{2\pi}\Re\left[F_{0}(\omega,t_{d})F_{G}^{*}(\omega,t_{d})\right]\boldsymbol{S}(\omega)\label{eq:beta0G}
\end{align}
are overlap matrix functions that include all the temporal dependence
of the magnetization decay. They are associated to the self-correlation
tensor function of the spin displacement $\left\langle \Delta\boldsymbol{x}(t_{1})\Delta\boldsymbol{x}(t_{2})\right\rangle $
and to the gradient modulation functions $f_{0}(t)$ and $f_{G}(t)$.
We consider modulation functions independent of the gradient direction.
Here the instantaneous displacement $\Delta\boldsymbol{x}(t)=\boldsymbol{x}(t)-\left\langle \boldsymbol{x}\right\rangle $
is the deviation from its average value. These overlap matrix functions
are also written in terms of their Fourier representations, where
the Fourier transform of the gradient modulation functions is
\begin{equation}
F(\omega,t_{d})=\int_{0}^{t_{d}}dt'\,e^{i\omega t'}f(t')
\end{equation}
and, using the Wiener-Khinchin theorem, the displacement power spectrum
matrix is 
\begin{equation}
S_{ij}(\omega)=\int_{\mathbb{-\infty}}^{\infty}dt'\,e^{i\omega t'}\left\langle \Delta x_{i}(t')\Delta x_{j}(0)\right\rangle ,
\end{equation}
assuming that the system is in the steady state.

The Fourier representation based on the displacement power spectrum
defines the dephasing of the spin signal from the overlap between
a filter function $\left|F(\omega,t)\right|^{2}$ that depends on
the gradient control and the frequencies modes distribution of the
diffusion process $\boldsymbol{S}(\omega)$ \citep{Stepisnik1993,Stepisnik2006,Lasic2006,Gore2010,Alvarez2013,Shemesh2013}.
Thus, the filter functions $\left|F(\omega,t)\right|^{2}$ can select
what frequency component of the displacement power spectrum affects
the dephasing.

Hence, the overlap matrix functions $\boldsymbol{\beta}(t)$ can be
tailored to probe the interaction of the diffusing spin-bearing particles
with the modulated gradients \citep{Alvarez2014,Alvarez2017}. The
matrices $\boldsymbol{\beta}^{GG}(t)$ and $\boldsymbol{\beta}^{00}(t)$
quantify the spin's interaction with the applied and background gradients
respectively, and $\boldsymbol{\beta}^{0G}(t)$ quantifies a cross-interaction
of the spins with the applied and background gradients \citep{Alvarez2017}.

The next non-Gaussian correction of Eq. (\ref{eq:cum_exp_phi}) is
the 4th order cumulant $\left\langle \left\langle \phi(t_{d})^{4}\right\rangle \right\rangle $.
This cumulant involves four time self-correlation tensor functions
such as $\left\langle \Delta\boldsymbol{x}(t_{1})\Delta\boldsymbol{x}(t_{2})\Delta\boldsymbol{x}(t_{3})\Delta\boldsymbol{x}(t_{4})\right\rangle $,
that can be simplified in terms of two times ones $\left\langle \Delta\boldsymbol{x}(t_{1})\Delta\boldsymbol{x}(t_{2})\right\rangle $
for Gaussian diffusion processes for the spin-bearing particle trajectory
$\boldsymbol{x}(t)$.

\subsection{Diffusion translation as a 3-dimensional Ornstein-Uhlembeck process\label{subsec:Diffusion-translation-OU}}

According to the \emph{internal gradient ensemble model}, the random
variable describing the diffusion motion is independent of the gradient
distribution. Thus, to obtain analytical results we model the spin-bearing
particles diffusion as a Gaussian and Markovian process according
to the assumptions discussed below. This diffusion model allow us
to simplify the high order self-correlation tensor function that appears
in Eq. (\ref{eq:cum_exp_phi}).

As typically assumed when modeling diffusion processes in restricted
compartments, we require the diffusion propagator to be stationary
when diffusion time tents to infinity. In one spatial dimension,
the only Gaussian, Markovian and stationary process is the Ornstein-Uhlenbeck
(OU) process \citep{Uhlenbeck_1930,Kuffer_2022}. In the three-dimensional
space, the OU process preserves these properties, therefore we consider
this model of diffusion that allows analytical calculations.

In a general restricted media, the displacement correlation function
along a given direction is given by 
\begin{equation}
\left\langle \Delta x(0)\Delta x(t)\right\rangle =D_{0}\sum_{k}b_{k}\tau_{k}e^{-|t|/\tau_{k}},\label{eq:general_correlation}
\end{equation}
where the coefficients $b_{k}$ and characteristic times $\tau_{k}$
account for the specific geometry of the compartment \citep{Wayne1966,Robertson1966,Stepisnik1993}.
Typically, the first of these exponential decays is the most significant,
dominating the spin signal evolution \citep{Alvarez2013,Shemesh2013,Zwick2020}.
The correlation function of the OU process is $\left\langle \Delta x(0)\Delta x(t)\right\rangle =D_{0}\tau_{c}e^{-|t|/\tau_{c}}$
with the characteristic correlation time $\tau_{c}$ \citep{Uhlenbeck_1930,Klauder1962,Stepisnik1999,Alvarez2013,Shemesh2013,Zwick2020,Capiglioni2021}.
Thus, modeling the diffusion with an OU process is equivalent to approximate
the correlation function of Eq. (\ref{eq:general_correlation}) to
the one given by its dominant term.

Based on this assumption, the effective restriction length $l_{c}$
of the microstructure compartment in which diffusion is taking place
can be defined as the variance of the stationary displacement distribution
of the OU process. This restriction length is defined by the correlation
time through the equation $l_{c}^{2}=D_{0}\tau_{c}$ \citep{Callaghan2011}.
The restriction length $l_{c}$ and the geometric size of the compartment
depend on its shape. For the example of cylinders oriented perpendicular
to the direction of the magnetic field gradient, a good approximation
is $l_{c}=0.37d$, where $d$ is the cylinder diameter \citep{Stepisnik1993,Stepisnik2006,Alvarez2013,Shemesh2013}.

\begin{figure}
\includegraphics[width=0.8\columnwidth]{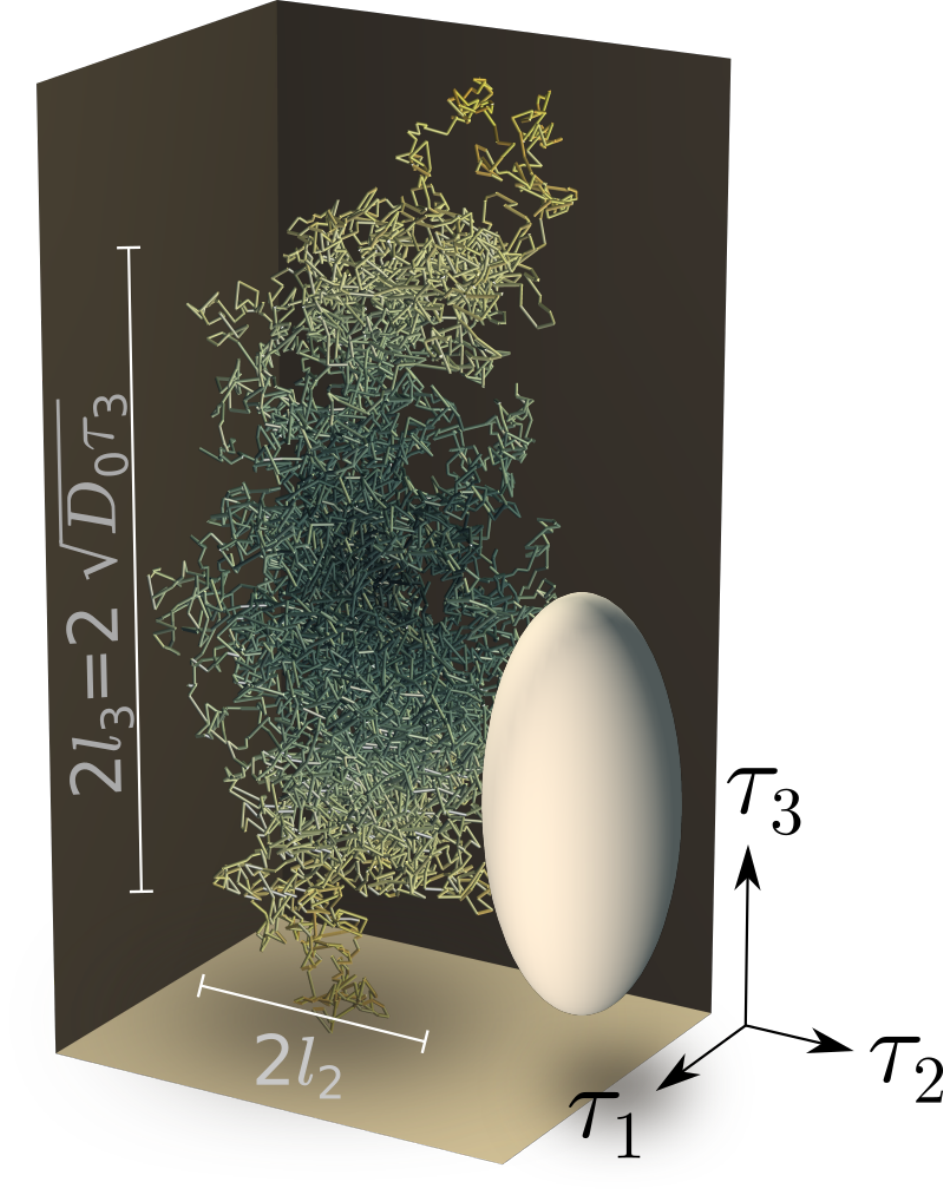}

\caption{\label{fig:RW3D}Schematic representation of anisotropic diffusion
due to anisotropic restriction lengths. The spin trajectory in space
is restricted, described by a OU stochastic process with restriction
lengths $l_{3}>l_{2}=l_{1}$, where $l_{i}^{2}=D_{0}\tau_{i}$. The
ellipsoid represents the corresponding correlation time tensor $\boldsymbol{\tau}_{c}$.}
\end{figure}

The three-dimensional version of the OU process \citep{Vatiwutipong2019}
accounts for anisotropic diffusion processes that are also typically
found in porous media and biological tissues \citep{Basser1994,Basser1994a,LeBihan2001,LeBihan2003,Le_Bihan_2012,Magdoom_2022}.
In order to generate a quantitative model that introduces the microstructure
morphology, we first consider that the free diffusion coefficient
is isotropic leading to the tensor $\boldsymbol{D}_{0}=D_{0}\boldsymbol{I}$.
The anisotropy of the diffusion process is thus only due to the compartment
morphology. This anisotropy is introduced from the different correlation
times along the different spatial directions in the displacement self-correlation
tensor 
\begin{equation}
\left\langle \Delta\boldsymbol{x}(t)\Delta\boldsymbol{x}(0)\right\rangle =D_{0}\boldsymbol{\tau}_{c}\exp\left(-\boldsymbol{\tau}_{c}^{-1}\left|t\right|\right).\label{eq:3d-displacement_correlation}
\end{equation}
Here the correlation-time tensor $\boldsymbol{\tau}_{c}$ has as eigenvalues
the diffusion correlation times $\tau_{i}$ along the principal axes
of the compartment geometry (Fig. \ref{fig:RW3D}). The Fourier transform
of displacement self-correlation tensor gives the three-dimensional
displacement power spectrum 
\begin{equation}
\boldsymbol{S}(\omega)=2D_{0}\left(\boldsymbol{\tau}_{c}^{-2}+\boldsymbol{I}\omega^{2}\right)^{-1}.\label{eq:spectral_dens}
\end{equation}
Thus the anisotropy in the displacement power spectrum tensor reflects
directly the anisotropy of the microstructure compartment morphology.
In the following, we calculate the overlap matrix functions of Eqs.
(\ref{eq:betaGG}-\ref{eq:beta0G}) based on this diffusion model.

\subsection{The IGDT expansion\label{subsec:The-IGDT-expansion}}

We consider the cumulant expansion up to the first non-Gaussian correction,
that is up to the 4th order cumulant of Eq. (\ref{eq:cum_exp_phi})
and thus assume the diffusion process for the spin-bearing particle
is driven by an OU process. The non-Gaussian effects increase with
the internal gradient variance $\left\langle \Delta G_{0}^{2}\right\rangle $,
where $\Delta\boldsymbol{G}_{0}=\boldsymbol{G}_{0}-\left\langle \boldsymbol{G}_{0}\right\rangle $.
It is thus expected the expansion of Eq. (\ref{eq:cum_exp_phi}) to
work well if $\left\Vert \boldsymbol{G}+\left\langle \boldsymbol{G}_{0}\right\rangle \right\Vert \gg\sqrt{\left\langle \Delta G_{0}^{2}\right\rangle }$.
Therefore, we rewrite Eq. (\ref{eq:cum_exp_phi}) in term of the moments
of the internal gradient distribution. In general, the cumulant expansion
(\ref{eq:cum_exp_phi}) converges slowly and irregularly with the
addition of higher order terms \citep{Jones2010}. Then, to improve
the convergence of the expansion up to the 4th order cumulant, we
omit the internal gradient distribution moments of order higher than
3. Under these assumptions, the magnetization decay is
\begin{align}
\ln M(t_{d})= & -\frac{1}{2}G_{i}\beta_{ij}^{GG}(t_{d})G_{j}-G_{i}\beta_{ij}^{0G}(t_{d})\left\langle \boldsymbol{G}_{0}\right\rangle _{j}\nonumber \\
 & -\frac{1}{2}\beta_{ij}^{00}(t_{d})\left\langle \boldsymbol{G}_{0}\boldsymbol{G}_{0}\right\rangle _{ij}\nonumber \\
 & +\frac{1}{2}G_{i}G_{l}\beta_{ij}^{0G}(t_{d})\beta_{lk}^{0G}(t_{d})\left\langle \Delta\boldsymbol{G}_{0}\Delta\boldsymbol{G}_{0}\right\rangle _{jk}\nonumber \\
 & +G_{i}\beta_{ij}^{0G}(t_{d})\beta_{lk}^{00}(t_{d})\left\langle \boldsymbol{G}_{0}\right\rangle _{l}\left\langle \Delta\boldsymbol{G}_{0}\Delta\boldsymbol{G}_{0}\right\rangle _{jk}\nonumber \\
 & +\frac{1}{2}\beta_{ij}^{00}(t_{d})\beta_{kl}^{00}(t_{d})\left\langle \boldsymbol{G}_{0}\right\rangle _{i}\left\langle \boldsymbol{G}_{0}\right\rangle _{l}\left\langle \Delta\boldsymbol{G}_{0}\Delta\boldsymbol{G}_{0}\right\rangle _{jk}\nonumber \\
 & +\frac{1}{2}G_{i}\beta_{ij}^{0G}(t_{d})\beta_{lk}^{00}(t_{d})\left\langle \Delta\boldsymbol{G}_{0}\Delta\boldsymbol{G}_{0}\Delta\boldsymbol{G}_{0}\right\rangle _{jkl}\nonumber \\
 & +\frac{1}{2}\beta_{ij}^{00}(t_{d})\beta_{lk}^{00}(t_{d})\left\langle \boldsymbol{G}_{0}\right\rangle _{i}\left\langle \Delta\boldsymbol{G}_{0}\Delta\boldsymbol{G}_{0}\Delta\boldsymbol{G}_{0}\right\rangle _{jkl}\nonumber \\
 & +\mathcal{O}\left[\left\langle \Delta\boldsymbol{G}_{0}^{4}\right\rangle \right],\label{eq:cum_expan}
\end{align}
where we used the Einstein notation again. The moments $\left\langle \boldsymbol{G}_{0}\boldsymbol{G}_{0}\cdots\right\rangle _{ij\cdots}$
and the central moments $\left\langle \Delta\boldsymbol{G}_{0}\Delta\boldsymbol{G}_{0}\cdots\right\rangle _{ij\cdots}$
denote the matrix elements of the IGDT. Notice that the average
is taken over the internal gradient distribution, expressing the cumulant
expansion of Eq. (\ref{eq:cum_exp_phi}) in terms of the IGDT of different
ranks.

The expansion of Eq. (\ref{eq:cum_expan}) involves several terms
that depend on the applied gradient strength $G$ and the IGDT of
different ranks, combined with the overlap integrals between gradient
modulation functions and the displacement self-correlations given
by the matrices $\boldsymbol{\beta}^{GG}(t)$, $\boldsymbol{\beta}^{00}(t)$
and $\boldsymbol{\beta}^{0G}(t)$. Equation (\ref{eq:cum_expan})
thus sets one of the main results of this article, defining what we
call the \emph{IGDT-expansion}. Table \ref{tab:terms_IGDT_expan}
summarize the notation we use in the following for all the terms of
Eq. (\ref{eq:cum_expan}).
\begin{table*}
\begin{tabular}{|c|c|c|c|}
\hline 
\textbf{IGDT groups} & \textbf{IGDT names} & \textbf{Definition} & \textbf{Description of the IGDT term}\tabularnewline
\hline 
$\chi_{\mathrm{app}}$ & $\chi_{G^{2}}$ & $-\frac{1}{2}G_{i}\beta_{ij}^{GG}(t)G_{j}$ & Applied gradient weighting\tabularnewline
\hline 
\multirow{6}{*}{$\chi_{\mathrm{bck}}$} & \multirow{2}{*}{$\chi_{0^{2}}$} & \multirow{2}{*}{$-\frac{1}{2}\beta_{ij}^{00}(t)\left\langle \boldsymbol{G}_{0}\boldsymbol{G}_{0}\right\rangle _{ij}$} & \tabularnewline
 &  &  & Trace-weighting of the internal gradient variance tensor\tabularnewline
\cline{2-4} \cline{3-4} \cline{4-4} 
 & \multirow{2}{*}{$\chi_{0^{2}\Delta^{2}}$} & \multirow{2}{*}{$\frac{1}{2}\beta_{ij}^{00}(t)\beta_{kl}^{00}(t)\left\langle \boldsymbol{G}_{0}\right\rangle _{i}\left\langle \boldsymbol{G}_{0}\right\rangle _{l}\left\langle \Delta\boldsymbol{G}_{0}\Delta\boldsymbol{G}_{0}\right\rangle _{jk}$} & Weighting of the internal gradient variance tensor\tabularnewline
 &  &  & in the mean internal gradient direction\tabularnewline
\cline{2-4} \cline{3-4} \cline{4-4} 
 & \multirow{2}{*}{$\chi_{0\Delta^{3}}$} & \multirow{2}{*}{$\frac{1}{2}\beta_{ij}^{00}(t)\beta_{kl}^{00}(t)\left\langle \boldsymbol{G}_{0}\right\rangle _{i}\left\langle \Delta\boldsymbol{G}_{0}\Delta\boldsymbol{G}_{0}\Delta\boldsymbol{G}_{0}\right\rangle _{jkl}$} & \tabularnewline
 &  &  & Weighting of the internal gradient distribution skewness\tabularnewline
\hline 
\multirow{7}{*}{$\chi_{\mathrm{odd-cross}}$} & \multirow{2}{*}{$\chi_{0G}$} & \multirow{2}{*}{$-G_{i}\beta_{ij}^{0G}(t)\left\langle \boldsymbol{G}_{0}\right\rangle _{j}$} & Weighting of the internal mean gradient\tabularnewline
 &  &  & projected into the applied gradient direction\tabularnewline
\cline{2-4} \cline{3-4} \cline{4-4} 
 & \multirow{3}{*}{$\chi_{0G\Delta^{2}}$} & \multirow{3}{*}{$G_{i}\beta_{ij}^{0G}(t)\beta_{kl}^{00}(t)\left\langle \boldsymbol{G}_{0}\right\rangle _{l}\left\langle \Delta\boldsymbol{G}_{0}\Delta\boldsymbol{G}_{0}\right\rangle _{jk}$} & Weighting of the internal gradient variance tensor\tabularnewline
 &  &  & projected in the applied and\tabularnewline
 &  &  & internal mean gradient directions\tabularnewline
\cline{2-4} \cline{3-4} \cline{4-4} 
 & \multirow{2}{*}{$\chi_{G\Delta^{3}}$} & \multirow{2}{*}{$\frac{1}{2}G_{i}\beta_{ij}^{0G}(t)\beta_{kl}^{00}(t)\left\langle \Delta\boldsymbol{G}_{0}\Delta\boldsymbol{G}_{0}\Delta\boldsymbol{G}_{0}\right\rangle _{jkl}$} & Cross-weighting of the internal gradient distribution\tabularnewline
 &  &  & skewness with the applied gradient\tabularnewline
\hline 
\multirow{2}{*}{$\chi_{\mathrm{even-cross}}$} & \multirow{2}{*}{$\chi_{G^{2}\Delta^{2}}$} & \multirow{2}{*}{$\frac{1}{2}G_{i}G_{l}\beta_{ij}^{0G}(t)\beta_{kl}^{0G}(t)\left\langle \Delta\boldsymbol{G}_{0}\Delta\boldsymbol{G}_{0}\right\rangle _{jk}$} & Internal gradient variance tensor\tabularnewline
 &  &  & weighted in the applied gradient direction\tabularnewline
\hline 
\end{tabular}

\caption{\label{tab:terms_IGDT_expan}Nomenclature for the different IGDT terms
of the cumulant expansion, their definitions and physical description.
The term descriptions in the last column give geometric interpretation
assuming isotropic diffusion, \emph{i.e.} overlap matrices $\boldsymbol{\beta}(t)$
proportional to the identity. The more general interpretation is based
on considering the overlap matrices as metric tensors that modify
the inner product of the gradient vectors. In this case the overlap
matrices $\boldsymbol{\beta}^{GG}(t)$ and $\boldsymbol{\beta}^{00}(t)$
are real, symmetric and positive-definite while $\boldsymbol{\beta}^{0G}(t)$
may be null, negative or positive-definite depending on the gradient
modulation symmetries. Given a non-singular symmetric matrix $\boldsymbol{\beta}$,
we can define an inner product $(\cdot,\cdot)$ between two vectors
$\boldsymbol{u}$ and $\boldsymbol{v}$ as $(\boldsymbol{u},\boldsymbol{v})=\boldsymbol{u}\cdot\boldsymbol{\beta}\cdot\boldsymbol{v}$.
Here $\boldsymbol{\beta}$ is the metric tensor and is a generalization
of the traditional Euclidean inner product $\boldsymbol{u}\cdot\boldsymbol{v}=\boldsymbol{u}\cdot\boldsymbol{I}\cdot\boldsymbol{v}$.
The definitions in the third column can be interpreted by this inner
product generalization with a metric defined by the anisotropy of
the diffusion process encoded by the $\boldsymbol{\beta}$ matrices.}
\end{table*}

The leading term $\chi_{G^{2}}=-\frac{1}{2}G_{i}\beta_{ij}^{GG}(t_{d})G_{j}$
only depends on the applied gradient modulation. The next term $\chi_{0G}=-G_{i}\beta_{ij}^{0G}(t_{d})\left\langle \boldsymbol{G}_{0}\right\rangle _{j}$
is a \emph{cross-term} involving the overlap integral of the displacement
power spectrum with the applied and internal gradient modulation filters.
The overlap matrix $\boldsymbol{\beta}^{0G}(t)$ can be positive,
negative or null depending on the gradient modulation shapes and symmetries,
unlike the attenuation matrices $\boldsymbol{\beta}^{GG}(t)$ and
$\boldsymbol{\beta}^{00}(t)$ that are always positive defined. The
magnitude of $\chi_{0G}$ depends on the angle between the applied
gradient and the internal gradient, therefore it can be controlled
by the applied gradient direction to probe the internal gradient direction.
Accordingly, its sign depends on the direction of $\boldsymbol{G}$.
The next term $\chi_{0^{2}}=-\frac{1}{2}\beta_{ij}^{00}(t_{d})\left\langle \boldsymbol{G}_{0}\boldsymbol{G}_{0}\right\rangle _{ij}$
is proportional to the mean square value of the background gradient,
but it is insensitive to its spatial orientation.

The next relevant term is $\chi_{G^{2}\Delta^{2}}=\frac{1}{2}G_{i}G_{l}\beta_{ij}^{0G}(t_{d})\beta_{kl}^{0G}(t_{d})\left\langle \Delta\boldsymbol{G}_{0}\Delta\boldsymbol{G}_{0}\right\rangle _{jk}$.
This is also a \emph{cross-term} similarly to $\chi_{0G}$, but it
involves the variance IGDT $\left\langle \Delta\boldsymbol{G}_{0}\Delta\boldsymbol{G}_{0}\right\rangle $.
This variance tensor provides information about the anisotropy of
the internal gradient distribution widths along the different spatial
directions, and thus depends on the media morphology \citep{Alvarez2017}.
Then, the term $\chi_{0^{2}\Delta^{2}}=\frac{1}{2}\beta_{ij}^{00}(t_{d})\beta_{kl}^{00}(t_{d})\left\langle \boldsymbol{G}_{0}\right\rangle _{i}\left\langle \boldsymbol{G}_{0}\right\rangle _{l}\left\langle \Delta\boldsymbol{G}_{0}\Delta\boldsymbol{G}_{0}\right\rangle _{jk}$
depends only on the internal gradient and contains information about
its distribution width. However, it is invariant under applied gradient
modulations, thus it does not provide information about internal gradient
anisotropy. The term $\chi_{0G\Delta^{2}}=G_{i}\beta_{ij}^{0G}(t_{d})\beta_{kl}^{00}(t_{d})\left\langle \boldsymbol{G}_{0}\right\rangle _{l}\left\langle \Delta\boldsymbol{G}_{0}\Delta\boldsymbol{G}_{0}\right\rangle _{jk}$
is a \emph{cross-term} as $\chi_{G^{2}\Delta^{2}}$ and $\chi_{0G}$,
and in this case its magnitude can be controlled by the applied gradient.
By suitable control of the symmetries of the gradient's modulations
it may vanish. Moreover, the sign of $\chi_{0G\Delta^{2}}$ changes
also like $\chi_{0G}$ when the direction of the applied gradient
is inverted. The last two terms $\chi_{G\Delta^{3}}=\frac{1}{2}G_{i}\beta_{ij}^{0G}(t_{d})\beta_{kl}^{00}(t_{d})\left\langle \Delta\boldsymbol{G}_{0}\Delta\boldsymbol{G}_{0}\Delta\boldsymbol{G}_{0}\right\rangle _{jkl}$
and $\chi_{0\Delta^{3}}=\frac{1}{2}\beta_{ij}^{00}(t_{d})\beta_{kl}^{00}(t_{d})\left\langle \boldsymbol{G}_{0}\right\rangle _{i}\left\langle \Delta\boldsymbol{G}_{0}\Delta\boldsymbol{G}_{0}\Delta\boldsymbol{G}_{0}\right\rangle _{jkl}$,
involve IGDT tensors of the 3th order that give information about
the skewness of the internal gradient distribution. They thus vanish
for symmetric distributions.

\section{Distilling and improving the IGDT expansion effects on the spin dephasing\label{sec:Improving-IGDT-eff}}

\subsection{Observing IGDT terms}

\begin{figure}
\includegraphics[width=1\columnwidth]{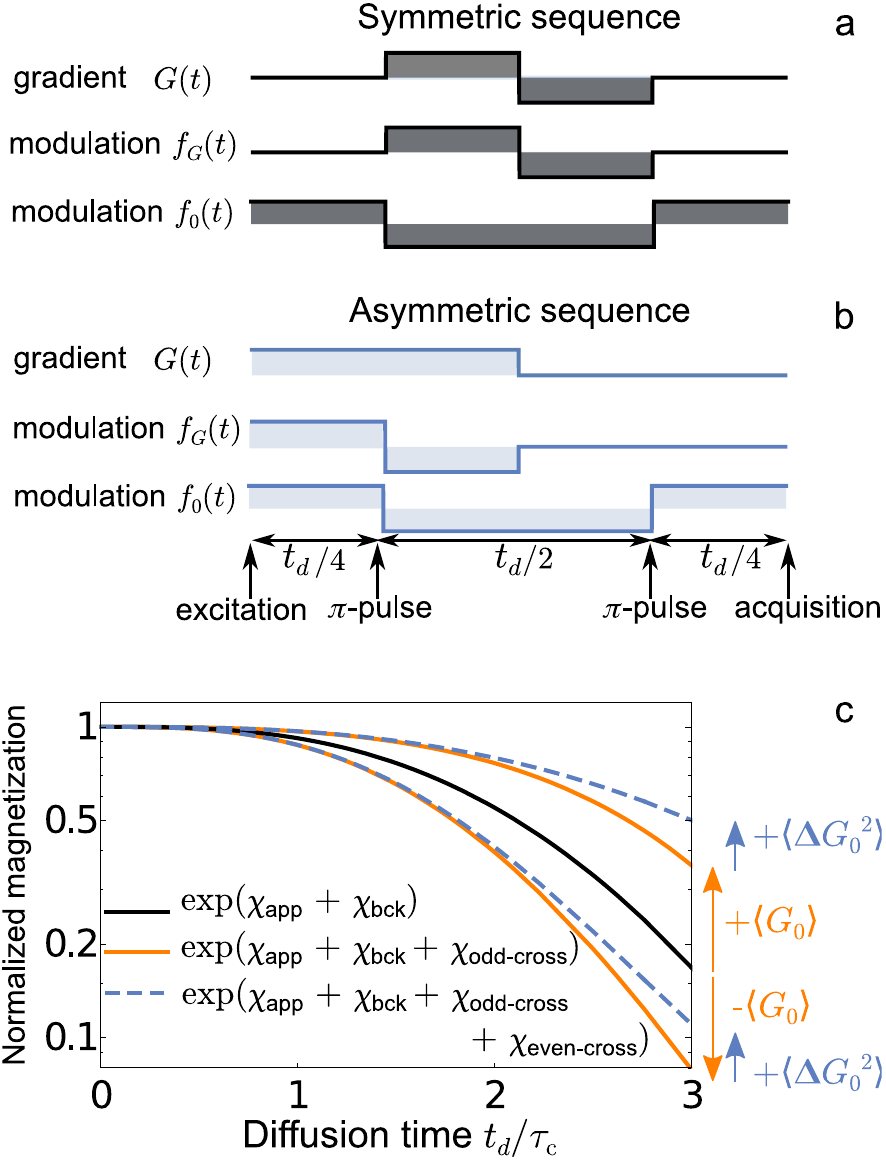}

\caption{\label{fig:esquema_signal}Schematic representation of a \emph{symmetric}
(a), an \emph{asymmetric sequence} (b) and the corresponding spin
signals behavior based on the contribution of the different IGDT terms
(c). The solid black line in (c) corresponds to the attenuation induced
by the pure applied (\emph{i}) and pure background gradients (\emph{ii})
terms derived from the symmetric sequence. The orange solid lines
add the \emph{odd cross-terms} contributions (\emph{iii}) to the signal
attenuation. Depending on the relative direction between the applied
and mean background gradient, this contribution is positive or negative.
The blue dashed lines in (c) are the signal decay derived from the
asymmetric sequence, that add \emph{the even cross-term} contribution,
which is proportional to the covariance tensor of the internal gradient
distribution. The \emph{even-cross terms} are always positive independent
on the applied gradient direction.}
\end{figure}

We aim at distilling the contribution of the different IGDT-expansion
terms in Eq. (\ref{eq:cum_expan}) and maximizing their effects on
the spin magnetization decay. We use MGSE sequences to avoid a phase
shift induced by the mean macroscopic inhomogeneity in the magnetization
signal according to Eq. (\ref{eq:MGSE_modulation_def}) \citep{Alvarez2014}.
We then identify four groups of terms in the IGDT-expansion $\ln M=\chi_{\mathrm{app}}+\chi_{\mathrm{bck}}+\chi_{\mathrm{odd-cross}}+\chi_{\mathrm{even-cross}}$
of Eq. (\ref{eq:cum_expan}) (see Table \ref{tab:terms_IGDT_expan}):
\begin{lyxlist}{00.00.0000}
\item [{(\emph{i})}] \emph{the pure applied} \emph{term} $\chi_{\mathrm{app}}=\chi_{G^{2}}$;
\item [{(\emph{ii})}] \emph{the pure background} \emph{terms}: $\chi_{\mathrm{bck}}=\chi_{0^{2}}+\chi_{0^{2}\Delta^{2}}+\chi_{0\Delta^{3}}$;
\item [{(\emph{iii})}] \emph{the odd cross-terms} $\chi_{\mathrm{odd-cross}}=\chi_{0G}+\chi_{0G\Delta^{2}}+\chi_{G\Delta^{3}}$;
\item [{(\emph{iv})}] and\emph{ the even cross-term} $\chi_{\mathrm{even-cross}}=\chi_{G^{2}\Delta^{2}}$.
\end{lyxlist}
These terms can be probed selectively by suitable design of the gradient
modulation sequence inspired on a previous proposal \citep{Alvarez2017}.
We can observe directly \emph{the pure applied} and \emph{background}
gradient terms (\emph{i} and \emph{ii}) $\chi_{\mathrm{app}}+\chi_{\mathrm{bck}}$,
by making null \emph{the cross-terms }contribution described in (\emph{iii})
and (\emph{iv}). \emph{The cross-terms }are null if the overlap matrix
$\boldsymbol{\beta}^{0G}(t_{d})=0$, by generating an odd function
for the product $f_{0}(t)f_{G}(t)$ with respect to the diffusion
time $t_{d}/2$. This is done by exploiting the symmetries of applied
and background gradient modulation to generate what we call the \emph{symmetric
sequence} as shown in Fig. \ref{fig:esquema_signal}a. Thus, the magnetization
attenuation factor for this sequence is $\ln M_{\mathrm{Sym}}=\chi_{\mathrm{app}}+\chi_{\mathrm{bck}}$.
The \emph{pure background} gradient effects (\emph{ii}) can be selectively
observed by a null applied gradient, \emph{i.e.} we obtain $\ln M_{\mathrm{Sym,}\boldsymbol{G}=0}=\chi_{\mathrm{bck}}$.
Then by subtracting its corresponding attenuation factor to the one
obtained by the \emph{symmetric sequence}, we can probe \emph{the
pure applied} gradient terms $\chi_{\mathrm{app}}=\ln M_{\mathrm{Sym}}-\ln M_{\mathrm{Sym,}\boldsymbol{G}=0}$.

We then increase the cross overlap matrix $\boldsymbol{\beta}^{0G}(t_{d})$
contribution while keeping invariant $\boldsymbol{\beta}^{GG}(t_{d})$
and $\boldsymbol{\beta}^{00}(t_{d})$. This is done by only relatively
shifting the applied and background gradient modulations functions,
creating what we call an \emph{asymmetric sequence }as shown in\emph{
}Fig. \ref{fig:esquema_signal}b\emph{.} Thus, by subtracting the
attenuation factors from the decaying signals between the \emph{symmetric}
and \emph{asymmetric sequences}, we can selectively observe the \emph{cross-terms}
contributions $\chi_{\mathrm{odd-cross}}+\chi_{\mathrm{even-cross}}=\ln M_{\mathrm{Asym}}-\ln M_{\mathrm{Sym}}$.
We then exploit the fact that the \emph{odd cross-terms} change their
sign by inverting the applied gradient direction, while \emph{the
even cross-terms} are invariant against this change. Again, by subtracting
or adding the attenuation factors derived from the inverted directions
of the applied gradient we can selectively probe \emph{the odd cross-terms}
\begin{equation}
\chi_{\mathrm{odd-cross}}=1/2\ln M_{\mathrm{Asym,+\boldsymbol{G}}}-1/2\ln M_{\mathrm{Asym,-\boldsymbol{G}}}\label{eq:odd-cross}
\end{equation}
 and \emph{the even cross-term}
\begin{multline}
\chi_{\mathrm{even-cross}}=1/2\ln M_{\mathrm{Asym,+\boldsymbol{G}}}+1/2\ln M_{\mathrm{Asym,-\boldsymbol{G}}}\\
\hphantom{1/2\ln M_{\mathrm{Asym,+\boldsymbol{G}}}+1/2aaaaaaa}-\ln M_{\mathrm{Sym}}.\label{eq:even-cross}
\end{multline}
Figure \ref{fig:esquema_signal}c shows a scheme manifesting the described
signal behaviors. 

For a 3-dimensional IGDT inference, we must repeat this strategy on
multiple non-collinear applied gradient directions in order to determine
the IGDT terms. Inspired on the methods for reconstructing the diffusion
tensor \citep{Basser1994,Basser1994a}, we proposed a technique to
extract $\left\langle \boldsymbol{G}_{0}\right\rangle $ and $\left\langle \Delta\boldsymbol{G}_{0}\Delta\boldsymbol{G}_{0}\right\rangle $
in Ref. \citep{Alvarez2017}. First, we need to determine the pure
applied gradient overlap matrix $\boldsymbol{\beta}^{GG}(t)$ to find
the correlation-time tensor $\boldsymbol{\tau_{c}}$ and the free
diffusion coefficient $D_{0}$. The matrix $\boldsymbol{\beta}^{GG}(t)$
has six independent elements. Therefore, we must measure \emph{the
pure applied} \emph{term} $\chi_{\mathrm{app}}$ at least for six
non-collinear applied gradient directions. As the attenuation terms
are noisy, it is also important to increase the number of non-collinear
gradient directions \citep{Mori2002,Caruyer_2013,Basser2002} to estimate
$\boldsymbol{\beta}^{GG}(t)$ statistically using a multivariate linear
regression. The correlation times in each direction and the free diffusion
coefficient can be estimated by fitting the estimated $\boldsymbol{\beta}^{GG}(t)$
based on our analytic model (\ref{eq:betaGG}).

Once these parameters are determined, we can calculate the other two
overlap matrices $\boldsymbol{\beta}^{0G}(t)$ and $\boldsymbol{\beta}^{00}(t)$
in term of them using Eqs. (\ref{eq:beta00}), (\ref{eq:beta0G})
and (\ref{eq:spectral_dens}). With the overlap matrix $\boldsymbol{\beta}^{0G}(t)$
reconstructed, we can estimate the IGDT $\left\langle \Delta\boldsymbol{G}_{0}\Delta\boldsymbol{G}_{0}\right\rangle $
and the mean internal gradient vector $\left\langle \boldsymbol{G}_{0}\right\rangle $
by measuring \emph{the even cross-term} $\chi_{\mathrm{even-cross}}$
and \emph{the odd cross-terms} $\chi_{\mathrm{odd-cross}}$ at several
applied gradient directions and using the multivariate linear regression.

\subsection{Sequence design to enhance the IGDT contributions\label{subsec:Sequence-design}}

The IGDT contributions may be weak. Due to the overlap matrix $\boldsymbol{\beta}^{0G}(t_{d})$
appears as a dyadic product in \emph{the cross-terms} $\chi_{G^{2}\Delta^{2}}$,
$\chi_{0G\Delta^{2}}$ and $\chi_{0^{2}\Delta^{2}}$, they become
more relevant than \emph{the pure applied} and \emph{pure background}
\emph{gradient} ones as the diffusion time increases. This can be
seen by considering the asymptotic behavior of the attenuation factors
$\beta(t_{d})\sim2c\gamma^{2}D_{0}\tau_{c}^{2}t_{d}$ and $\beta(t_{d})\sim2c\gamma^{2}D_{0}t_{d}^{3}$,
for the restricted ($t_{d}\gg\tau_{c}$) and free ($t_{d}\ll\tau_{c}$)
diffusion limits, respectively. Here $c$ is a dimensionless coefficient
that depends on the specific gradient modulation waveform. The first
order terms that are linearly proportional to the overlap matrices
$\boldsymbol{\beta}(t_{d})$ are the most significant on the attenuation
factor up to the following approximated time scale 
\begin{equation}
t_{G_{tot}}=\left[c\gamma^{2}D_{0}\left(\boldsymbol{G}+\left\langle \boldsymbol{G}_{0}\right\rangle \right)^{2}\right]^{-1/3},\label{eq:charac_time_1st_order_(free)}
\end{equation}
 for the free diffusion regime and 
\begin{equation}
t_{G_{tot}}=\left[c\gamma^{2}D_{0}\left(\boldsymbol{G}+\left\langle \boldsymbol{G}_{0}\right\rangle \right)^{2}\tau_{c}^{2}\right]^{-1},\label{eq:charac_time_1st_order_(rest)}
\end{equation}
for the restricted diffusion regime. Here, $t_{G_{tot}}$ is the dephasing
time associated to the magnetization $M(t_{d})=\exp\left\{ -\frac{1}{2}\beta(t_{d})\left(\boldsymbol{G}+\left\langle \boldsymbol{G}_{0}\right\rangle \right)^{2}\right\} $
decay to $1/e$, of the spin-bearing particles diffusing in a gradient
$\boldsymbol{G}_{tot}=\boldsymbol{G}+\left\langle \boldsymbol{G}_{0}\right\rangle $
in the respective diffusion regimes. Beyond these times, the higher
order IGDT terms become more relevant.

The previous consideration are general for a cumulant expansion. In
the following, we consider control strategies to enhance the IGDT
terms contribution to the signal decay, while minimizing the loss
of signal due to the pure interaction with the applied gradient. We
need to enhance the terms that include the cross-overlap matrix $\boldsymbol{\beta}^{0G}(t)$
in the IGDT expansion of Eq. (\ref{eq:cum_expan}). We thus look for
conditions on the overlap matrices $\boldsymbol{\beta}^{GG}(t_{d})$
and $\boldsymbol{\beta}^{0G}(t_{d})$ that increase \emph{the cross-terms
}contribution with respect to \emph{the pure applied} one. 

We find that requesting the matrix difference $|\boldsymbol{\beta}^{0G}(t_{d})|-\boldsymbol{\beta}^{GG}(t_{d})$
to be positive-definite, enhance the IGDT terms contributions. Here,
we define the modulus matrix $|\boldsymbol{\beta}|$ as the matrix
whose eigenvalues are the modulus of the eigenvalues of $\boldsymbol{\beta}$.
As the principal directions of all considered overlap matrices are
defined by the ones of the correlation-time tensor $\boldsymbol{\tau}_{c}$,
the positive-definite condition for $|\boldsymbol{\beta}^{0G}(t_{d})|-\boldsymbol{\beta}^{GG}(t_{d})$
is
\begin{equation}
|\beta_{i}^{0G}(t_{d})|>\beta_{i}^{GG}(t_{d})\ \ \ i=1,2,3,\label{eq:beta_0G > beta_GG}
\end{equation}
where $\beta_{i}^{0G}(t_{d})$ and $\beta_{i}^{GG}(t_{d})$ are the
eigenvalues of the corresponding overlap matrices. A derivation for
this condition to enhance the IGDT contributions is found in the Appendix
\ref{sec:General-demonstration}. Notice that from Eq. (\ref{eq:cum_expan}),
if the vector $\boldsymbol{\beta}^{0G}(t_{d})\left\langle \boldsymbol{G}_{0}\right\rangle $
is orthogonal to the applied gradient, the second term vanishes independently
on the relation (\ref{eq:beta_0G > beta_GG}). Then, the expression
of Eq. (\ref{eq:beta_0G > beta_GG}) is a necessary but not sufficient
condition.

\begin{figure}
\includegraphics[width=1\columnwidth]{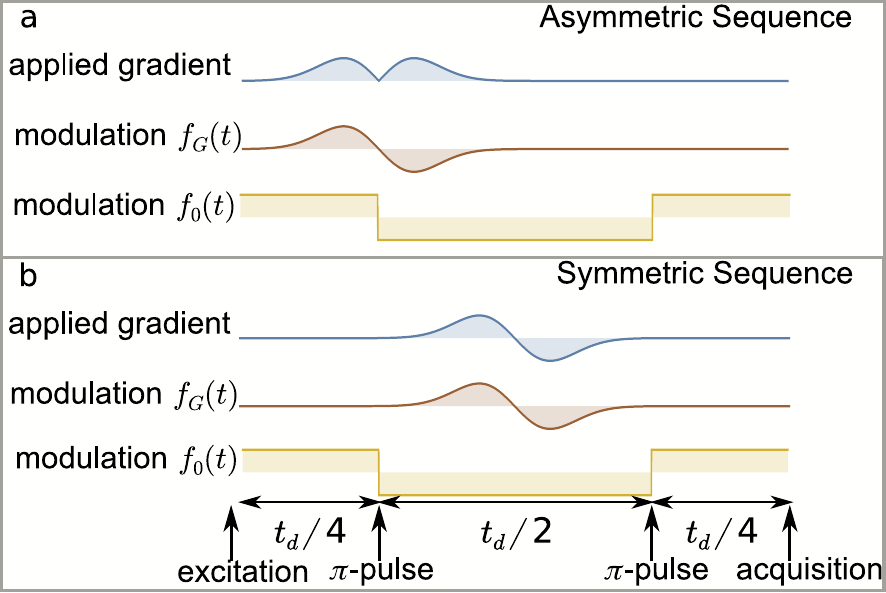}

\caption{\label{fig:modulation_scheme}Suitable design of \emph{Symmetric }and\emph{
Asymmetric Sequences} of modulated gradients to probe the IGDTs. The
sequences involve an initial excitation pulse followed by an applied
gradient modulation defined by $f_{G}(t)$ given by Eq. (\ref{eq:fg_modulation}).
Together with two RF $\pi$-pulses at times $t=1/4\,t_{d}$ and $t=3/4\,t_{d}$
to modulate the background gradient as $f_{0}(t)$, determines the
Gaussian derivative gradient spin-echo (GDGSE) sequence. This diffusion
weighting block is then followed by an acquisition block. (a) \emph{Asymmetric
Sequence} and (b) \emph{Symmetric Sequence}. In both sequences, the
internal gradient modulation $f_{0}(t)$ corresponds to the one determined
by the two RF pulses conforming a CPMG sequence that refocus the decoherence
due to susceptibility-induced magnetic field inhomogenities. The applied
gradient modulation is proportional to the first derivative of the
Gaussian function as shown in Eq. (\ref{eq:fg_modulation}). The spin
dephasing due to interaction with the positive part of the applied
gradient modulation is then refocused by the interaction with the
negative part of the modulation. The modulation profile $f_{G}(t)$
in the \emph{Asymmetric Sequence} is obtained by applying the gradient
profile shown with the blue line in panel (b), as the RF $\pi$-pulses
effectively change the sign of the phase evolution of the spins. The
time interval in which the modulation $f_{G}(t)$ is significantly
different from zero is lower than $1/2\,t_{d}$.}
\end{figure}

The pure background gradient contribution is enhanced by minimizing
the number of RF pulses, as its corresponding filter function lower
the frequency of its dominant peak, thus maximizing the overlap with
the displacement spectral density $\boldsymbol{S}(\omega)$ which
is centered at zero frequency. \emph{The cross-term} contributions
are enhanced by increasing the applied gradient strength {[}see Eq.
(\ref{eq:cum_expan}){]}, however this also enhance the pure applied
gradient term. In order to avoid the latter effect, we increase the
modulation frequency of the applied gradient to reduce the overlap
of the applied gradient filter function with the spectral density,
thus reducing its dephasing effects. This can be done by concentrating
the temporal variations of $f_{G}(t)$ in a small interval as can
be seen in Fig. \ref{fig:modulation_scheme}.

If the modulation frequency of the applied gradient becomes larger
than the one of the background gradient modulation, the cross-filter
overlap $\Re\left[F_{0}(\omega,TE)F_{G}^{*}(\omega,TE)\right]$ in
Eq. (\ref{eq:beta0G}) also is reduced. With the aim of increasing
\emph{the cross-terms} dephasing effect, we thus reduce the total
duration of the applied gradient modulation to broaden its high-frequency
peak and allow the overlap with the background gradient filter component
at lower frequencies. This cross-filter overlap can be further maximized
by synchronizing properly the sign changes of the applied and background
gradient modulations to make a constructive interference between them
as described in details in the Appendix \ref{sec:Optimizing-the-cross-filter's}.

In order to attain the condition of Eq. (\ref{eq:beta_0G > beta_GG}),
we reduce the effective interaction time of the spins with the applied
gradient, compared to the times they interact with the cross interference
of the applied and background gradients. The effective interaction
time of the background and applied gradients are the diffusion times
during which the spins interact with each of these gradients, 
\begin{align}
T_{0} & =\int_{0}^{t_{d}}dt\,f_{0}^{2}(t)\label{eq:T0=00003DTE}
\end{align}
and 
\begin{align}
T_{G} & =\int_{0}^{t_{d}}dt\,f_{G}^{2}(t),\label{eq:TG}
\end{align}
respectively. Similarly, the effective interaction time of \emph{the
cross-term} is the diffusion time that generates the cross interference
between the applied and background gradients,

\begin{align}
T_{0G} & =\int_{0}^{t_{d}}dt\,f_{0}(t)f_{G}(t).\label{eq:T0G}
\end{align}
Therefore, to achieve the condition (\ref{eq:beta_0G > beta_GG}),
we need $T_{0G}>T_{G}$. This latter condition can be seen more directly
at the restricted diffusion limit, where we can approximate the overlap
matrices as 
\begin{align}
\boldsymbol{\beta}_{rest}^{GG}(t_{d}) & \approx\boldsymbol{S}(0)\gamma^{2}\int_{\mathbb{-\infty}}^{\infty}\frac{d\omega}{2\pi}\left|F_{G}(\omega,t_{d})\right|^{2}\nonumber \\
 & \approx\boldsymbol{S}(0)\gamma^{2}\int_{0}^{t_{d}}dt\,f_{G}^{2}(t)\nonumber \\
 & \approx\boldsymbol{S}(0)\gamma^{2}T_{G}
\end{align}
 and 
\begin{align}
\boldsymbol{\text{\ensuremath{\beta}}}_{rest}^{0G}(t_{d}) & \approx\boldsymbol{S}(0)\gamma^{2}\int_{\mathbb{-\infty}}^{\infty}\frac{d\omega}{2\pi}\Re\left[F_{0}(\omega,t_{d})F_{G}^{*}(\omega,t_{d})\right]\nonumber \\
 & \approx\boldsymbol{S}(0)\gamma^{2}\int_{0}^{t_{d}}dt\,f_{0}(t)f_{G}(t)\nonumber \\
 & \approx\boldsymbol{S}(0)\gamma^{2}T_{0G},
\end{align}
being both proportional to the corresponding effective interaction
times, as the displacement spectral density can be factored out from
the overlap integral. A general demonstration is found in the Appendix
\ref{sec:General-demonstration}.

Applied gradients can be arbitrary modulated in general. However,
the background gradient is only modulated by applying the RF $\pi$-pulses
that invert the spin phase evolution, which is encoded in the modulation
function $f_{0}(t)$ that switches between $+1$ and $-1$ at every
time a $\pi$-pulses is applied. To enhance \emph{the cross-term}
effect with respect to \emph{the pure applied term,} we thus propose
the implementation of smooth modulations for the applied gradient
that satisfy the condition $|f_{G}(t)|^{2}<|f_{G}(t)|$, and a synchronization
of the sign changes of $f_{0}(t)$ and $f_{G}(t)$ such that $f_{0}(t)f_{G}(t)=|f_{G}(t)|$
(Fig. \ref{fig:modulation_scheme}).

To get a MGSE sequence, the applied gradient modulation $f_{G}(t)$
must cross zero at least once. The condition $f_{0}(t)f_{G}(t)=|f_{G}(t)|$
can be attained by setting a smooth MGSE with an applied gradient
modulation with a single refocusing echo, whose zero crossing time
matches the instant of time where the RF $\pi$-pulse is applied to
modulate the backgrounmd gradient function $f_{0}(t)$. To avoid further
sign inversions in $f_{0}(t)f_{G}(t)$, the applied gradient modulation
$f_{G}(t)$ must vanish every time that a $\pi$-pulse is applied.
Figure \ref{fig:modulation_scheme}a shows an example of such a modulation.
Thus, all of these conditions define a general \emph{asymmetric sequence
}that enhance the IGDT \emph{cross-term} contributions\emph{.}

To probe directly the IGDT terms proportional to $\boldsymbol{\beta}^{GG}(t_{d})$
and $\boldsymbol{\beta}^{00}(t_{d})$, we must cancel \emph{the cross-term}
dephasing effect in Eq. (\ref{eq:cum_expan}), while keeping invariant
\emph{the pure terms}. We thus make $\boldsymbol{\beta}^{0G}(t_{d})=0$
by setting $f_{G}(t)$ an odd function with respect to its middle
time and $f_{0}(t)$ an even function again with respect to its middle
time. Then, the cross-filter vanishes by matching the modulation centers
of $f_{G}(t)$ and $f_{0}(t)$. Figure \ref{fig:modulation_scheme}b
shows an example of such modulation sequence. The pure overlap matrices
$\boldsymbol{\beta}^{GG}(t_{d})$ and $\boldsymbol{\beta}^{00}(t_{d})$
thus remain invariant, because temporal shifts of the corresponding
modulation functions does not alter the squared modulus of their Fourier
transform. This shifted modulation thus define a general \emph{symmetric
sequence.}

Finally, to maximize the background gradient dephasing effects with
the overlap integrals $\boldsymbol{\beta}^{00}(t_{d})$ and $\left|\boldsymbol{\beta}^{0G}(t_{d})\right|$,
we have to modulate the background gradient with a low frequency to
maximize their overlap with the displacement spectral density $\boldsymbol{S}(\omega)$.
We thus need a modulation sequence with the minimal possible number
of pulses. As we require that $f_{0}(t)$ be an even function, we
choose a Carr-Purcell-Meiboom-Gill (CPMG) sequence \citep{Carr1954,Meiboom1958}
with two pulses for its modulation. The RF $\pi$-pulses are thus
applied at $t=1/4\,t_{d}$ and $t=3/4\,t_{d}$ respectively (Fig.
\ref{fig:modulation_scheme}).

In the Appendix \ref{sec:General-demonstration}, we show that the
relation of Eq. (\ref{eq:beta_0G > beta_GG}) is fulfilled assuming
the sequence conditions derived in this section for an arbitrary diffusion
regime.

In summary, to enhance \emph{the cross-term} effects, the background
gradient modulation $f_{0}(t)$ has to be an even function with respect
to $t_{d}/2$ with a minimal number of pulses. The applied gradient
modulation has to be an MGSE sequence with a smooth modulation function
such that $|f_{G}(t)|\leq1$, with a single refocusing echo matching
the zero crossing time with a $\pi$-pulse that modulate $f_{0}(t)$.
The modulation function $f_{G}(t)$ has to be an odd function with
respect to its middle time, \emph{i.e.} the zero crossing time, and
should vanish also at every $\pi$-pulse that modulate $f_{0}(t)$.

\subsection{Paradigmatic example: Gaussian derivative modulation}

To follow the above sequence design, we propose here a paradigmatic
MGSE sequence with applied gradient modulations derived from a Gaussian
function derivative. We call it the Gaussian derivative gradient spin-echo
(GDGSE) modulation that allow us to obtain analytical calculations.
The GDGSE sequence consists on a modulation function $f_{G}(t)$ determined
by the first derivative of a Gaussian function centered at $t_{d}/2$
with a standard deviation given by a fraction $\alpha\,t_{d}$ of
the total diffusion time $t_{d}$ (see Fig. \ref{fig:modulation_scheme}b)
\begin{equation}
f_{G}(t)=\frac{\sqrt{e}}{\alpha t_{d}}(t-t_{d}/2)e^{-\frac{(t-t_{d}/2)^{2}}{2\alpha^{2}t_{d}^{2}}}.\label{eq:fg_modulation}
\end{equation}
Here, $\alpha\ll1$ is a coefficient that controls the Gaussian modulation
width.

The modulation function is normalized such that it satisfies $\mathrm{max}[f_{G}(t)]=1$,
so as the maximum gradient strength $G$ is achieved. While the GDGSE
allows us to obtain analytical expressions for the dephasing, its
modulation is not finite along the time domain. As the applied gradient
modulation decays exponentially for times $t\gg\alpha t_{d}$, we
set the parameter $\alpha\ll1$ to make negligible its modulation
amplitude at every RF $\pi$-pulse and outside the considered diffusion
time interval $t_{d}$. We thus consider negligible the modulation
amplitude at the RF pulses $f_{G}(1/4\,t_{d})=f_{G}(3/4\,t_{d})\text{\ensuremath{\approx0}}$
and outside the diffusion probing time. For example, by setting $\alpha=1/15$,
the modulation amplitude at the RF pulses and outside the probing
time is $\lesssim10^{-3}$. This requirement for $\alpha$ also provides
a large frequency for the applied gradient modulation as required
in Sec. \ref{subsec:Sequence-design} and, at the same time, reduces
the effective interaction time of the applied gradient of Eq. (\ref{eq:TG}).

The \emph{asymmetric sequence} is obtained by shifting the modulation
function $f_{G}(t)$ by $t_{d}/4$ to the left, as seen in Fig. \ref{fig:modulation_scheme}a.
To ensure a significant integral overlap between the applied and background
gradient filters $\Re\left[F_{0}(\omega,t_{d})F_{G}^{*}(\omega,t_{d})\right]$,
the coefficient $\alpha$ must not be too small. We found that $1/30\lesssim\alpha\lesssim1/10$
works well enough. To find a proper $\alpha$, we need to compare
the GDGSE filter maximum $|F_{G}(1/(\alpha t_{d}),t_{d})|^{2}$ {[}see
Eq. (\ref{eq: filtro FG}) in Appendix \ref{sec:Filters-and-overlap}{]}
with the GDGSE filter value at the modulation frequency of the background
gradient $2\pi/t_{d}$. For $\alpha=1/15$, for example, $|F_{G}(2\pi/t_{d},t_{d})|^{2}/|F_{G}(1/(\alpha t_{d}),t_{d})|^{2}=0.4$.

\subsection{Overlap matrix weightings for diffusion asymptotic limits}

Here we consider two physically relevant asymptotic regimes to evaluate
the overlap matrix weighting on the IGDT expansion terms, the free
diffusion ($t_{d}/\tau_{c}\ll1$) and the restricted diffusion ($t_{d}/\tau_{c}\gg1$)
limits. The general expressions for the overlap integrals of Eqs.
(\ref{eq:betaGG}-\ref{eq:beta0G}), can be found in the Appendix
\ref{sec:Filters-and-overlap}. As expected at the free diffusion
regime, the behavior of the three overlap functions is cubic on the
evolution time. If the condition $t_{d}/\tau_{c}\ll1$ is fulfilled
in all spatial directions, the diffusion can be assumed isotropic.
The overlap integrals are then scalars as all principal directions
are equivalent, where 
\begin{equation}
\beta_{free}^{GG}(t_{d})=2e\sqrt{\pi}\gamma^{2}D_{0}\alpha^{3}t_{d}^{3}
\end{equation}
and 
\begin{equation}
\beta_{free}^{00}(t_{d})=\frac{\gamma^{2}D_{0}}{24}t_{d}^{3}.
\end{equation}
 As we consider $\alpha\ll1$ to make $f_{G}(t+t_{d}/4)\approx0$
at the edge of the interval $[0,t_{d}/2]$, the free diffusion limit
of the cross overlap function gives the cross overlap 
\begin{equation}
\beta_{free}^{0G}(t_{d})=\sqrt{\frac{\pi e}{2}}\left(1/\alpha-4\sqrt{\frac{2}{\pi}}\right)\gamma^{2}D_{0}\alpha^{3}t_{d}^{3}.
\end{equation}

Instead, at the restricted diffusion limit, all the overlap matrices
are lineal as a function of the diffusion time 
\begin{equation}
\boldsymbol{\beta}_{rest}^{GG}(t_{d})=e\sqrt{\pi}\gamma^{2}D_{0}\boldsymbol{\tau_{c}}^{2}\alpha t_{d},
\end{equation}
 
\begin{equation}
\boldsymbol{\beta}_{rest}^{00}(t_{d})=2\gamma^{2}D_{0}\boldsymbol{\tau_{c}}^{2}(t_{d}-5\boldsymbol{\tau_{c}}).
\end{equation}
 Again, as $\alpha\ll1$, the restricted diffusion limit of the cross
overlap matrix can be approximated as 
\begin{equation}
\boldsymbol{\beta}_{rest}^{0G}(t_{d})=4\sqrt{e}\gamma^{2}D_{0}\boldsymbol{\tau_{c}}^{2}\alpha t_{d}.
\end{equation}

The advantage of using the smooth gradient modulations to obtain the
positive-definite condition for the matrix $|\boldsymbol{\beta}^{0G}(t_{d})|-\boldsymbol{\beta}^{GG}(t_{d})$
compared with sharp ones that switch the gradient sign between $\left\{ -1,+1\right\} $,
can be seen in these asymptotic behaviors. The ratios between the
overlap matrices for $\alpha\ll1$ are $\boldsymbol{\beta}_{free}^{0G}(t_{d})/\boldsymbol{\beta}_{free}^{GG}(t_{d})=\sqrt{\frac{e}{2}}\left(1/\alpha-4\sqrt{\frac{2}{\pi}}\right)/(2e)>1$
in the free diffusion limit and $\boldsymbol{\beta}_{rest}^{0G}(t_{d})/\boldsymbol{\beta}_{rest}^{GG}(t_{d})=4\sqrt{e}/(e\sqrt{\pi})>1$
in the restricted one. On the other hand, a sharp modulation that
only switches the gradient sign between $\left\{ -1,+1\right\} $,
can attain, at the most, the same effective interaction times for
the spin interaction with the applied gradient and with the cross-interference
between the applied and background gradients. Then, it is not possible
to minimize the interaction effects with the applied gradient without
also reducing the cross-interaction effects in the same ratio, and
thus no enhancement of the cross IGDT terms is found.

\section{The IGDT expansion in typical brain tissue conditions\label{sec:IGDT-in-brain}}

In this section we evaluate the IGDT expansion assuming typical values
of the free diffusion coefficient and correlation times in brain tissues.
The diffusion coefficient typically belongs to the range starting
from $0.8\,\mathrm{\mu m^{2}/ms}$ to $2.2\,\mathrm{\mu m^{2}/ms}$
for human brain \citep{Mulkern_1999,Maier_2001,Sener_2001} and $3\,\mathrm{\mu m^{2}/ms}$
for free water at $37\,\mathrm{{^\circ}C}$ \citep{Tofts_2000}. Brain
tissue microstruture sizes range for example for the axon radius,
typically between the interval $0.3\,\mathrm{\mu m}$-$3\,\mathrm{\mu m}$
in human brain \citep{Mikelberg1989,Barazany_2009,Huang2020}. Assuming
cylindrical geometries with a restriction radius between this range,
the expected diffusion correlation time $\tau_{c}$ ranges from $10^{-2}\,\mathrm{ms}$
to $10\,\mathrm{ms}$, based on the Fick-Einstein equation $l_{c}^{2}=D_{0}\tau_{c}$.
\begin{figure*}[t]
\includegraphics[width=0.8\textwidth]{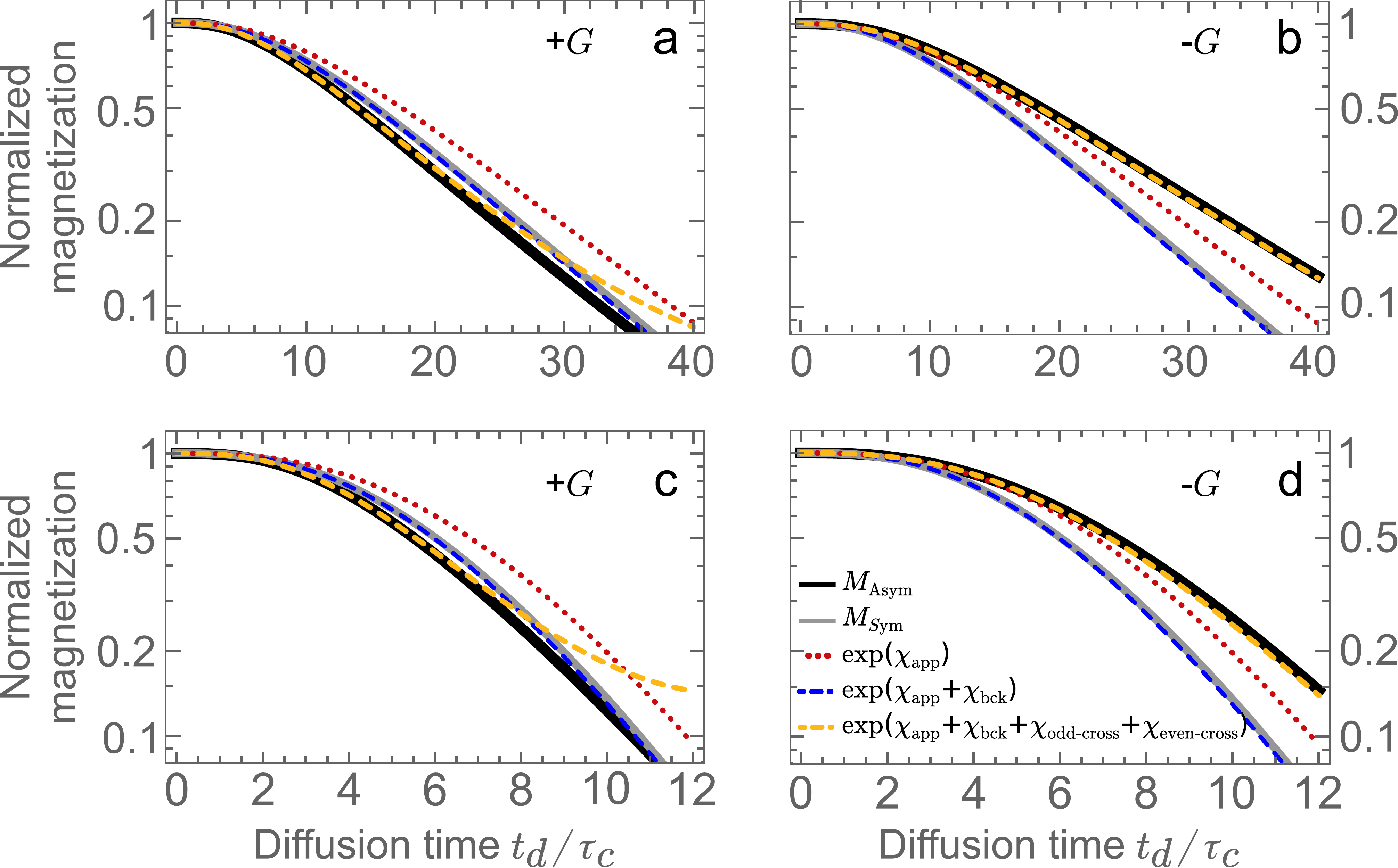}

\caption{\label{fig:igdt_expan}Contribution of different terms of the IGDT
expansion. The solid black lines correspond to the magnetization signal
produced by the \emph{asymmetric sequence}, while the gray ones corresponds
to the signal produced by the \emph{symmetric sequence}. These signals
are calculated numerically by averaging the magnetization signal with
a Gaussian distribution for the background gradient $G_{0}$ with
a standard deviation $\Delta G_{0}=0.15|G|$, and mean value $\left\langle G_{0}\right\rangle =0.1|G|$
in panels (a, c) and with opposite sign in panels (b, d). The red
dotted line corresponds to \emph{the pure applied gradient} contribution
($\chi_{\mathrm{app}}=\chi_{G^{2}}$), the blue dashed line includes\emph{
}also\emph{ the pure background gradient} contribution ($\chi_{\mathrm{bck}}=\chi_{0^{2}}+\chi_{0^{2}\Delta^{2}}$).
The yellow dashed lines includes \emph{the cross-terms} contributions
$\chi_{\mathrm{odd-cross}}=\chi_{0G}+\chi_{0G\Delta^{2}}$ and $\chi_{\mathrm{even-cross}}=\chi_{G^{2}\Delta^{2}}$.
We consider the free diffusion coefficient $D_{0}=2.2\,\mathrm{\mu m^{2}/ms}$,
a correlation time $\tau_{c}=2\,\mathrm{ms}$ in panels (a, b) and
$\tau_{c}=5\,\mathrm{ms}$ in panels (c, d) corresponding to restriction
lengths of $2\text{\ensuremath{\mu}m}$ and $3.3\text{\ensuremath{\mu\mathrm{m}}}$,
and a giromagnetic factor of $\gamma=267\,\mathrm{mT^{-1}ms^{-1}}$.
In panel (a) and (b) the applied gradient is $G=600\,\mathrm{mT/m}$
and in panels (c) and (d) $G=400\,\mathrm{mT/m}$.}
\end{figure*}
\begin{figure*}
\includegraphics[width=0.8\textwidth]{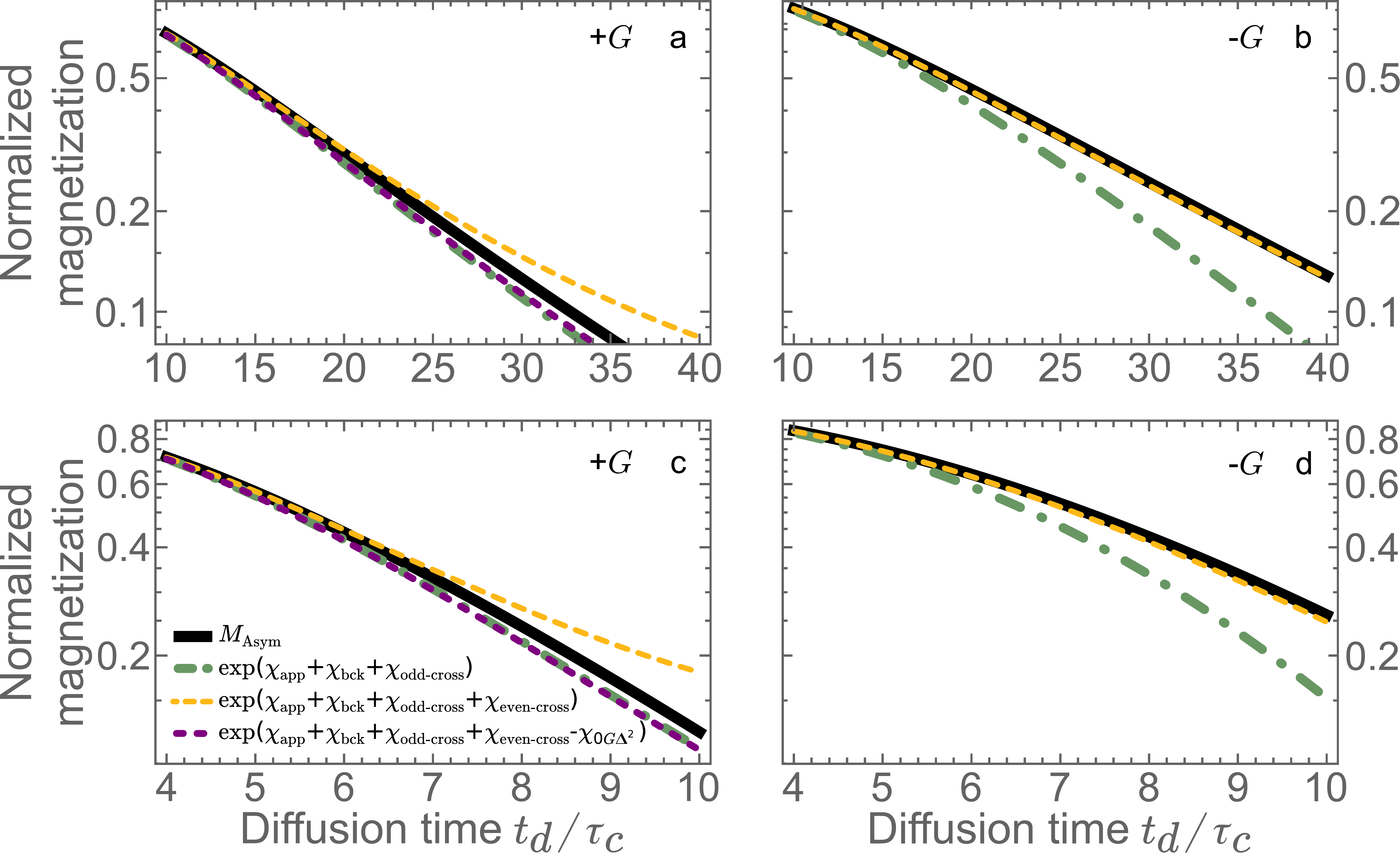}

\caption{\label{fig:igdt_zoom}Contribution of different terms of the IGDT
expansion to the magnetization signal produced by the \emph{asymmetric
sequence} (solid black lines). The green dot-dashed lines include
\emph{the pure applied} ($\chi_{\mathrm{app}}=\chi_{G^{2}}$),\emph{
pure background }($\chi_{\mathrm{bck}}=\chi_{0^{2}}+\chi_{0^{2}\Delta^{2}}$)
and\emph{ odd cross-terms} contribution ($\chi_{\mathrm{odd-cross}}=\chi_{0G}+\chi_{0G\Delta^{2}}$).
The yellow dashed lines includes also \emph{the} \emph{even cross-term}
contribution $\chi_{\mathrm{even-cross}}=\chi_{G^{2}\Delta^{2}}$.
The purple dashed line consider all the terms in the yellow curve
subtracting \emph{the odd-cross term} $\chi_{0G\Delta^{2}}=G\beta_{0G}(t_{d})\beta_{00}(t_{d})\left\langle G_{0}\right\rangle \left\langle \Delta G_{0}^{2}\right\rangle $.
As in Fig. \ref{fig:igdt_expan}, we consider the free diffusion coefficient
$D_{0}=2.2\,\mathrm{\mu m^{2}/ms}$, a correlation time $\tau_{c}=2\,\mathrm{ms}$
in panels (a, b) and $\tau_{c}=5\,\mathrm{ms}$ in panels (c, d),
and a giromagnetic factor $\gamma=267\,\mathrm{mT^{-1}ms^{-1}}$.
In panel (a) and (b) the applied gradient is $G=600\,\mathrm{mT/m}$
and in panels (c) and (d) $G=400\,\mathrm{mT/m}$.}
\end{figure*}

The internal magnetic field variations are of order of $10^{-6}B_{0}$
\citep{Han2011,Chen_2013,Xu_2017}. Intra and extra-axonal internal
gradients in white matter show mono-disperse internal gradient distribution
centered at zero and with a distribution width that is maximum in
a direction perpendicular to the axon direction \citep{Han2011,Chen_2013,Xu_2017,Alvarez2017}.
This was also demonstrated with capillary tube phantoms mimicking
the axon morphology giving internal gradient width of the order of
$1\,\mathrm{mT/m}$ on a $1\,\mathrm{T}$ magnet \citep{Han2011}.
Numerical calculations of internal gradients in axon bundles using
the AxonPacking open-source software \citep{Mingasson_2017} and a
Finite Perturber Method (FPM) \citep{Pathak_2008,Han2011}, predict
internal gradient widths on the range of 20-200 mT/m in a 9.4T magnet
\citep{Fajardo}. As internal gradients scale with the tissue compartmentalization
characteristic length $l_{c}$, internal gradient variations can be
estimated by $\Delta G_{0}\approx10^{-6}B_{0}/l_{c}$. Then, $\Delta G_{0}\approx1000\,\mathrm{mT/m}$
can be expected in brain tissues, consistently with observations in
numerical simulations \citep{Fajardo}. However, the highest internal
gradient strengths are typically concentrated only in small spatial
regions, so that the molecular movement averages them, reducing their
values effectively, thus making the standard deviation of internal
gradient distribution smaller.

To evaluate the main features of the IGDT-expansion, we focus on a
1-dimensional analysis associated with the diffusion process in one
of the principal axis of the correlation time tensor $\boldsymbol{\tau}_{c}$.
The analysis is equivalent for the three principal axes, in which
the only relevant aspect is how it scales with the corresponding $\text{\ensuremath{\tau_{c}}}$.
According to Eq. (\ref{eq:cum_expan}), the IGDT-expansion then reduces
to
\begin{align}
\ln M(t_{d})= & -\frac{1}{2}G^{2}\beta_{GG}(t_{d})-G\beta_{0G}(t_{d})\left\langle G_{0}\right\rangle \nonumber \\
 & -\frac{1}{2}\beta_{00}(t_{d})\left\langle G_{0}^{2}\right\rangle \nonumber \\
 & +\frac{1}{2}G^{2}\beta_{0G}^{2}(t_{d})\left\langle \Delta G_{0}^{2}\right\rangle \nonumber \\
 & +G\beta_{0G}(t_{d})\beta_{00}(t_{d})\left\langle G_{0}\right\rangle \left\langle \Delta G_{0}^{2}\right\rangle \nonumber \\
 & +\frac{1}{2}\beta_{00}^{2}(t_{d})\left\langle G_{0}\right\rangle ^{2}\left\langle \Delta G_{0}^{2}\right\rangle +\mathcal{O}\left[\left\langle \Delta G_{0}^{3}\right\rangle \right].\label{eq:1D_IGDT_expan}
\end{align}

To model the diffusion process, we consider an OU process in one spatial
dimension and the gradient ensemble description of Sec. \ref{subsec:Internal-Gradient-ensemble}.
The IGDT-expansion better approximates the real signal if the applied
gradient is stronger than the standard deviation of internal gradient
distribution $G\gg\Delta G_{0}$. Here we assume a Gaussian distribution
for the internal gradients with mean $\left\langle G_{0}\right\rangle =0.1G$
and standard deviation $\Delta G_{0}=0.15G$ for the internal gradients.
Based on the reported values of internal magnetic field variations
in brain tissue \citep{Xu_2017}, we consider for our simulations
two cases for the relation between the applied and internal gradients
$G=600\,\mathrm{mT/m}$, $\left\langle G_{0}\right\rangle =60\,\mathrm{mT/m}$
and $G=400\,\mathrm{mT/m}$, and $\left\langle G_{0}\right\rangle =40\,\mathrm{mT/m}$.
The internal gradient distribution widths are thus $\Delta G_{0}=90\:\mathrm{mT/m}$
and $\Delta G_{0}=60\,\mathrm{mT/m}$, respectively.

Based on Eq. (\ref{eq:magnetization_def_1}), we first calculate the
magnetization average of the spin-bearing particles diffusing in presence
of a given internal gradient strength $G_{0}$. The thermal diffusion
follows the OU Gaussian process as described in Sec. \ref{subsec:Diffusion-translation-OU}
for the spin's random phase. The resulting magnetization average thus
only depends on the second cumulant of the spin phase 
\begin{equation}
M(t_{d},G_{0})=e^{-\frac{1}{2}\left\langle \left\langle \phi(t_{d},G_{0})^{2}\right\rangle \right\rangle },\label{eq:M(G0)}
\end{equation}
with $\left\langle \left\langle \phi(t_{d},G_{0})^{2}\right\rangle \right\rangle =G^{2}\beta_{GG}(t_{d})+2G_{0}G\beta_{0G}(t_{d})+G_{0}^{2}\beta_{00}(t_{d})$.
Notice that this second cumulant is equal to the one in Eq. (\ref{eq:cum_2})
for a given $G_{0}$ value, \emph{i.e.} without yet considering the
internal gradient distribution. The overlap functions $\beta_{GG}(t_{d})$,
$\beta_{00}(t_{d})$ and $\beta_{0G}(t_{d})$ are given by the eigenvalues
corresponding to the considered principal axes of the attenuation
matrices of Eqs. (\ref{eq:betaGG}-\ref{eq:beta0G}). Then, we calculate
the total magnetization decay by averaging Eq. (\ref{eq:M(G0)}) over
the internal gradient distribution.

The magnetization signal for the\emph{ asymmetric sequence} described
in Fig. \ref{fig:modulation_scheme}a includes the \emph{cross-term}
in the gradient average
\begin{align}
M_{\mathrm{Asym}}(t_{d})= & e^{-\frac{1}{2}G^{2}\beta_{GG}(t_{d})}\nonumber \\
 & \times\left\langle e^{-G_{0}G\beta_{0G}(t_{d})}e^{-\frac{1}{2}G_{0}^{2}\beta_{00}(t_{d})}\right\rangle .\label{eq:M_asym}
\end{align}
Instead, the cross overlap integral $\beta_{0G}(t_{d})$ vanishes
for the \emph{symmetric sequence} described in Fig. \ref{fig:modulation_scheme}b,
and thus the magnetization signal is
\begin{equation}
M_{\mathrm{Sym}}(t_{d})=e^{-\frac{1}{2}G^{2}\beta_{GG}(t_{d})}\left\langle e^{-\frac{1}{2}G_{0}^{2}\beta_{00}(t_{d})}\right\rangle .
\end{equation}

Typical relaxation times $T_{2}$ of white and gray matter in brain
tissue are on the order of $\sim50-130\,\mathrm{ms}$ \citep{Bailes_1982,Holland1986,Wansapura_1999}.
We thus evaluate the effects of the different groups of terms in the
IGDT-expansion before this time scale to be able to predict the reliability
to measure them. Figure \ref{fig:igdt_expan} shows two relevant experimental
situations, a restricted (Figs. \ref{fig:igdt_expan}a,b) and a free
diffusion (Figs. \ref{fig:igdt_expan}c,d) regime. For illustration,
we assumed the free diffusion coefficient $D_{0}=2.2\,\mathrm{\mu m^{2}/ms}$
in Fig. \ref{fig:igdt_expan}. It is still unknown the free diffusion
coefficient in brain tissue, as typically only the apparent diffusion
coefficient is measured. If the free diffusion coefficient is lower
than the assumed one, the magnetization decay becomes slower and thus
more diffusion time is required for probing the IGDT effects. For
example, if the free diffusion coefficient reduces to $D_{0}\text{\ensuremath{\approx}}0.7\,\mathrm{\mu m^{2}/ms}$,
the behavior of the curves of Fig. \ref{fig:igdt_expan} would decay
to 10\% of the initial magnetization at 100 ms instead of 40 ms in
the cases of panels a and b; and at 20 ms instead of 12 ms in the
cases of panels c and d. Diffusion times that allow probing the IGDT
effects are thus still accessible for the typical $T_{2}$ values
of white and gray matter in brain even in this scenario.

We consider applied gradient modulations with opposite gradient directions
to show the behavior of the \emph{odd} and \emph{even cross terms}
in the cumulant expansion. In both regimes, the magnetization of the
\emph{symmetric sequence} is well described by \emph{the} \emph{pure
applied} $\chi_{\mathrm{app}}=-\frac{1}{2}G_{i}\beta_{ij}^{GG}(t_{d})G_{j}$
and \emph{pure background} \emph{gradient term}s $\chi_{\mathrm{bck}}=\chi_{0^{2}}+\chi_{0^{2}\Delta^{2}}+\chi_{0\Delta^{3}}$
of the IGDT-expansion of Eq. (\ref{eq:1D_IGDT_expan}) (see table
\ref{tab:terms_IGDT_expan}). Notice that the \emph{pure background}
\emph{gradient terms} increase the magnetization decay with respect
to the decay produced by only considering a \emph{pure applied gradient
term}. This is a consequence of the fact that the \emph{total background}
\emph{gradient term} is negative as the first term $\chi_{0^{2}}=-\frac{1}{2}\beta_{00}(t_{d})\left\langle G_{0}^{2}\right\rangle $
is larger than the next term $\chi_{0^{2}\Delta^{2}}=\frac{1}{2}\beta_{00}^{2}(t_{d})\left\langle G_{0}\right\rangle ^{2}\left\langle \Delta G_{0}^{2}\right\rangle $
with opposite sign. As we discussed in Sec. \ref{subsec:The-IGDT-expansion},
the lower order terms in the cumulant expansion with lower power law
exponents of overlap matrices are dominant at short times $t_{d}<t_{G_{tot}}$,
where $t_{G_{tot}}$ is given by Eqs. (\ref{eq:charac_time_1st_order_(free)})
and (\ref{eq:charac_time_1st_order_(rest)}), depending on the diffusion
regime. At longer times $t_{d}>t_{G_{tot}}$, the higher order terms
become relevant and they might then reduce the overall decay. By reducing
the effective interaction time of the applied gradient with respect
to the one of the background gradient, as done in the sequence of
Fig. \ref{fig:modulation_scheme}, also makes the \emph{pure background}
\emph{gradient terms} more significant with respect to the pure applied
terms.

Then, by including \emph{the cross-terms} $\chi_{odd-cross}+\chi_{even-cross}=\chi_{0G}+\chi_{0G\Delta^{2}}+\chi_{G^{2}\Delta^{2}}$
of the expansion (\ref{eq:1D_IGDT_expan}) (see Table \ref{tab:terms_IGDT_expan}),
we reproduce the magnetization decay generated by the \emph{asymmetric
sequence} (Figs. \ref{fig:igdt_expan}b,d). In Figs. \ref{fig:igdt_expan}b,d
there is a better convergence than in Figs. \ref{fig:igdt_expan}a,c,
where the convergence is poor for $t_{d}>25\tau_{c}$ and $t_{d}>7\tau_{c}$,
respectively. The performance difference of the cumulant expansion
is due to the relative direction between the applied and average background
gradient that change the sign of\emph{ the odd cross-terms}.

We now analyze this effect on the signal decay. The first two \emph{odd
cross-terms} $\chi_{0G}=-G\beta_{0G}(t_{d})\left\langle G_{0}\right\rangle $
and $\chi_{0G\Delta^{2}}=G\beta_{0G}(t_{d})\beta_{00}(t_{d})\left\langle G_{0}\right\rangle \left\langle \Delta G_{0}^{2}\right\rangle $
that we consider here, have opposite signs. Then, the behavior of
their overall decay contribution at short times is the opposite of
the one at longer times. \emph{The even cross-term} $\chi_{G^{2}\Delta^{2}}=\frac{1}{2}G^{2}\beta_{0G}^{2}(t_{d})\left\langle \Delta G_{0}^{2}\right\rangle $
is always positive and it thus always reduce the magnetization decay.
As it contains a quadratic order on the attenuation overlap function,
$\beta_{0G}^{2}(t_{d})$ becomes relevant at long times $t_{d}>t_{G_{tot}}$
(see Eqs. \ref{eq:charac_time_1st_order_(free)} and \ref{eq:charac_time_1st_order_(rest)}).
This is seen at times longer than $\sim15\tau_{c}$ in Fig. \ref{fig:igdt_zoom}b,
and at times longer than $\sim6\tau$ in Fig. \ref{fig:igdt_zoom}d,
when the green dash-dotted curve starts to deviate from the exact
decay (black line), and the dashed yellow line continues reproducing
the expected behavior.

In Fig. \ref{fig:igdt_expan}a and \ref{fig:igdt_expan}c, the applied
gradient has the same sign as the background gradient $\left\langle G_{0}\right\rangle $.
\emph{The} \emph{pure background} \emph{gradient term }$\chi_{0^{2}}$
and \emph{the odd cross-term} $\chi_{0G}$ thus contribute by increasing
the magnetization decay at short times, as both terms are negative.
At longer times the behavior changes,\emph{ }as \emph{the odd cross-term}
$\chi_{0G\Delta^{2}}$ and \emph{the even cross-term $\chi_{G^{2}\Delta^{2}}$
}now tend to reduce the magnetization decay because they are positive.
In these cases, the dashed yellow line diverges from the expected
signal behavior at times longer than $25\tau_{c}$ in Fig. \ref{fig:igdt_zoom}a
and at times longer than $7\tau_{c}$ in Fig. \ref{fig:igdt_zoom}c.
At these times the terms $\chi_{0G\Delta^{2}}$ and \emph{$\chi_{G^{2}\Delta^{2}}$}
become more relevant. This shows the slow convergence of the IGDT
expansion when the applied gradient is in the same direction of the
mean background gradient. In this case, we obtain a better approximation
by including only \emph{the even cross-term} \emph{$\chi_{G^{2}\Delta^{2}}$
}such as\emph{
\[
\ln M(t_{d})=\chi_{\mathrm{app}}+\chi_{\mathrm{bck}}+\chi_{0G}+\chi_{G^{2}\Delta^{2}},
\]
} and excluding the \emph{odd cross-term} $\chi_{0G\Delta^{2}}$ or
\emph{vice versa} \emph{
\[
\ln M(t_{d})=\chi_{\mathrm{app}}+\chi_{\mathrm{bck}}+\chi_{0G}+\chi_{0G\Delta^{2}}.
\]
} Figures \ref{fig:igdt_zoom}a and c show that these expansion's
approximations (purple dashed line and green dot-dashed respectively)
reproduce better the predicted curve (black line). 
\begin{figure}[b]
\includegraphics[width=0.98\columnwidth]{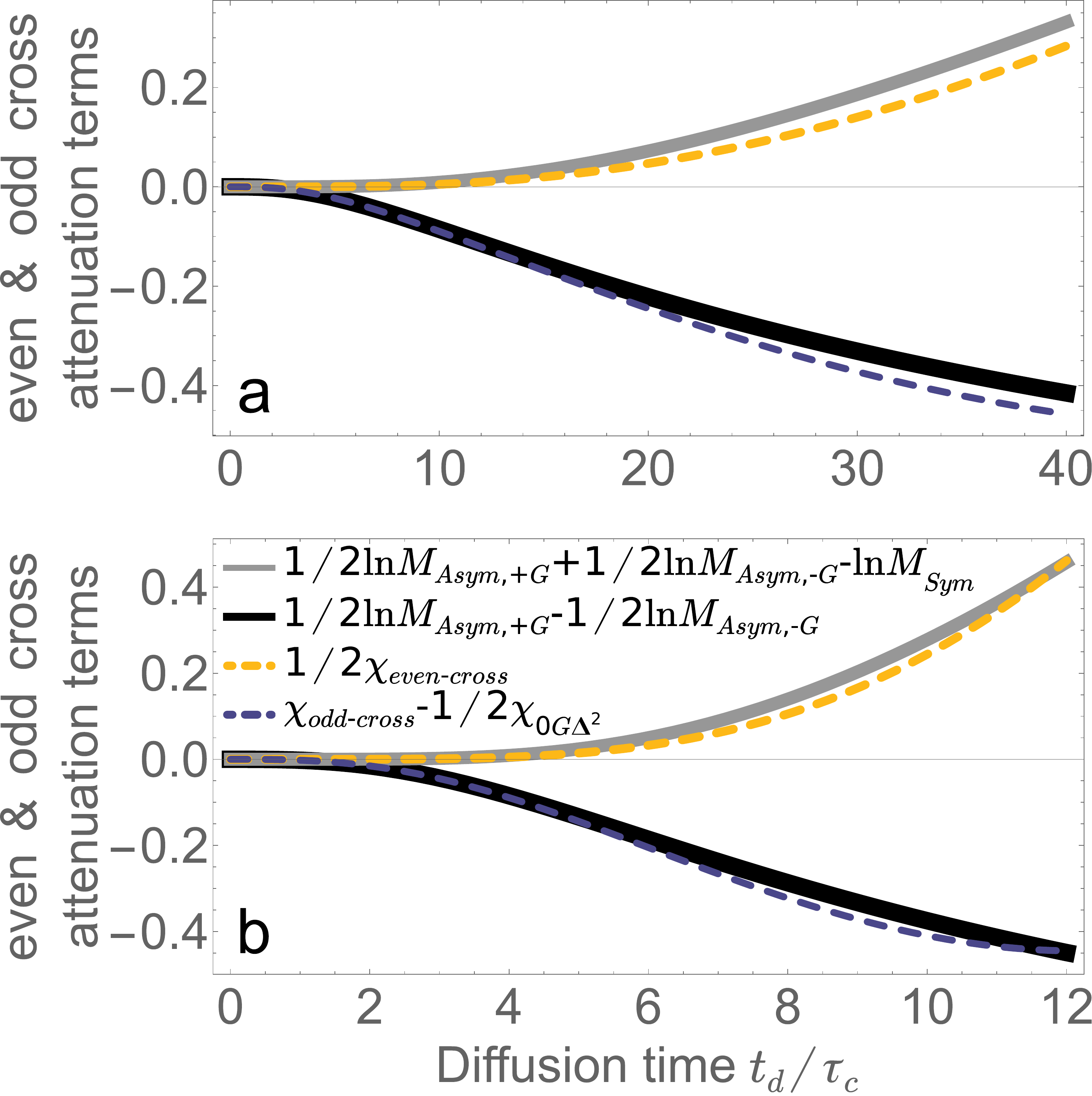}\caption{\label{fig:Even_and_odd}Approximated \emph{even }and \emph{odd cross-term}
attenuations when the applied gradient has the same direction as the
background gradient. The gray and black solid lines gives the approximated
\emph{even }and \emph{the odd cross-terms} following Eqs. (\ref{eq:eve-approx})
and (\ref{eq:odd-approx}) respectively, calculated from the mathematical
operations between the magnetization signal of the \emph{symmetric}
and \emph{asymmetric sequences} with the two applied gradient directions.
These magnetization signals are calculated numerically by averaging
the magnetization signal with a Gaussian distribution for the background
gradient $G_{0}$ with a standard deviation $\Delta G_{0}=0.15|G|$,
and mean value $\left\langle G_{0}\right\rangle =0.1|G|$. The yellow
and purple dashed lines show the corresponding\emph{ even cross-term}
$1/2\chi_{\mathrm{even-cross}}=1/2\chi_{G^{2}\Delta^{2}}$ and \emph{the
odd cross-term} $\chi_{0G}+1/2\chi_{0G\Delta^{2}}=\chi_{\mathrm{odd-cross}}-1/2\chi_{0G\Delta^{2}}$
of the left-hand-side of Eqs. (\ref{eq:eve-approx}) and (\ref{eq:odd-approx})
respectively, derived from the expansion of Eq. (\ref{eq:1D_IGDT_expan}).
As in Fig. \ref{fig:igdt_expan}, we consider the free diffusion coefficient
$D_{0}=2.2\,\mathrm{\mu m^{2}/ms}$, a correlation time $\tau_{c}=2\,\mathrm{ms}$
in panel (a) and $\tau_{c}=5\,\mathrm{ms}$ in panel (b), and a giromagnetic
factor $\gamma=267\,\mathrm{mT^{-1}ms^{-1}}$. In panel (a) the applied
gradient is $G=600\,\mathrm{mT/m}$ and in panel (b) $G=400\,\mathrm{mT/m}$.}
\end{figure}

In Fig. \ref{fig:igdt_expan}b and \ref{fig:igdt_expan}d, the applied
gradient has the opposite sign of the average background gradient
$\left\langle G_{0}\right\rangle $. In this case, the signal decay
is reduced at short times since the leading \emph{odd cross-term}
$\chi_{0G}$ reduces the magnetization decay. At long times $t_{d}>t_{G_{tot}}$
(see Eqs. \ref{eq:charac_time_1st_order_(free)} and \ref{eq:charac_time_1st_order_(rest)}),
\emph{the even cross-term $\chi_{G^{2}\Delta^{2}}$} and the \emph{odd
cross-term} $\chi_{0G\Delta^{2}}$ have opposite behavior allowing
a better convergence (Fig. \ref{fig:igdt_zoom}b and d). As a result,
the contrast ratio $M_{\mathrm{Asym}}(t_{d})/M_{\mathrm{Sym}}(t_{d})$
between the magnetization of the \emph{asymmetric} and \emph{symmetric
sequences} is maximized when applied and background gradient are in
the opposite direction compared with the case when they are in the
same direction.

As shown in Fig. \ref{fig:igdt_zoom}a and c, when the applied gradient
has the same direction as the background gradient, the green and purple
curves are good approximation and indistinguishable between them.
In the green curve, we exclude \emph{the even-cross term} $\chi_{G^{2}\Delta^{2}}$
and in the purple curve, \emph{the odd-cross term} $\chi_{0G\Delta^{2}}$
from the expansion. As it is arbitrary which of them is excluded,
we propose under this condition to approximate \emph{the even-cross
term} from 
\begin{multline}
1/2\chi_{\mathrm{even-cross}}\text{\ensuremath{\approx}}1/2\ln M_{\mathrm{Asym,+\boldsymbol{G}}}+1/2\ln M_{\mathrm{Asym,-\boldsymbol{G}}}\\
\hphantom{1/2\ln M_{\mathrm{Asym,+\boldsymbol{G}}}+1/2aaaaaaa}-\ln M_{\mathrm{Sym}},\label{eq:eve-approx}
\end{multline}
rather than from Eq. (\ref{eq:even-cross}) by excluding the term
$\chi_{\mathrm{even-cross}}$ from $\ln M_{\mathrm{Asym,+\boldsymbol{G}}}$.
To determine \emph{the odd cross-term}, then use the approximation
\begin{multline}
\chi_{0G}+1/2\chi_{0G\Delta^{2}}=\chi_{\mathrm{odd-cross}}-1/2\chi_{0G\Delta^{2}}\\
\text{\ensuremath{\approx}}1/2\ln M_{\mathrm{Asym,+\boldsymbol{G}}}-1/2\ln M_{\mathrm{Asym,-\boldsymbol{G}}},\label{eq:odd-approx}
\end{multline}
rather than Eq. (\ref{eq:odd-cross}) by excluding the term $\chi_{0G\Delta^{2}}$
from $\ln M_{\mathrm{Asym,+\boldsymbol{G}}}$. Figure \ref{fig:Even_and_odd}
shows the validity of these assumptions.

\section{Validity of the internal gradient ensemble model\label{sec:scop-limitations}}

\begin{figure*}
\includegraphics[width=0.9\textwidth]{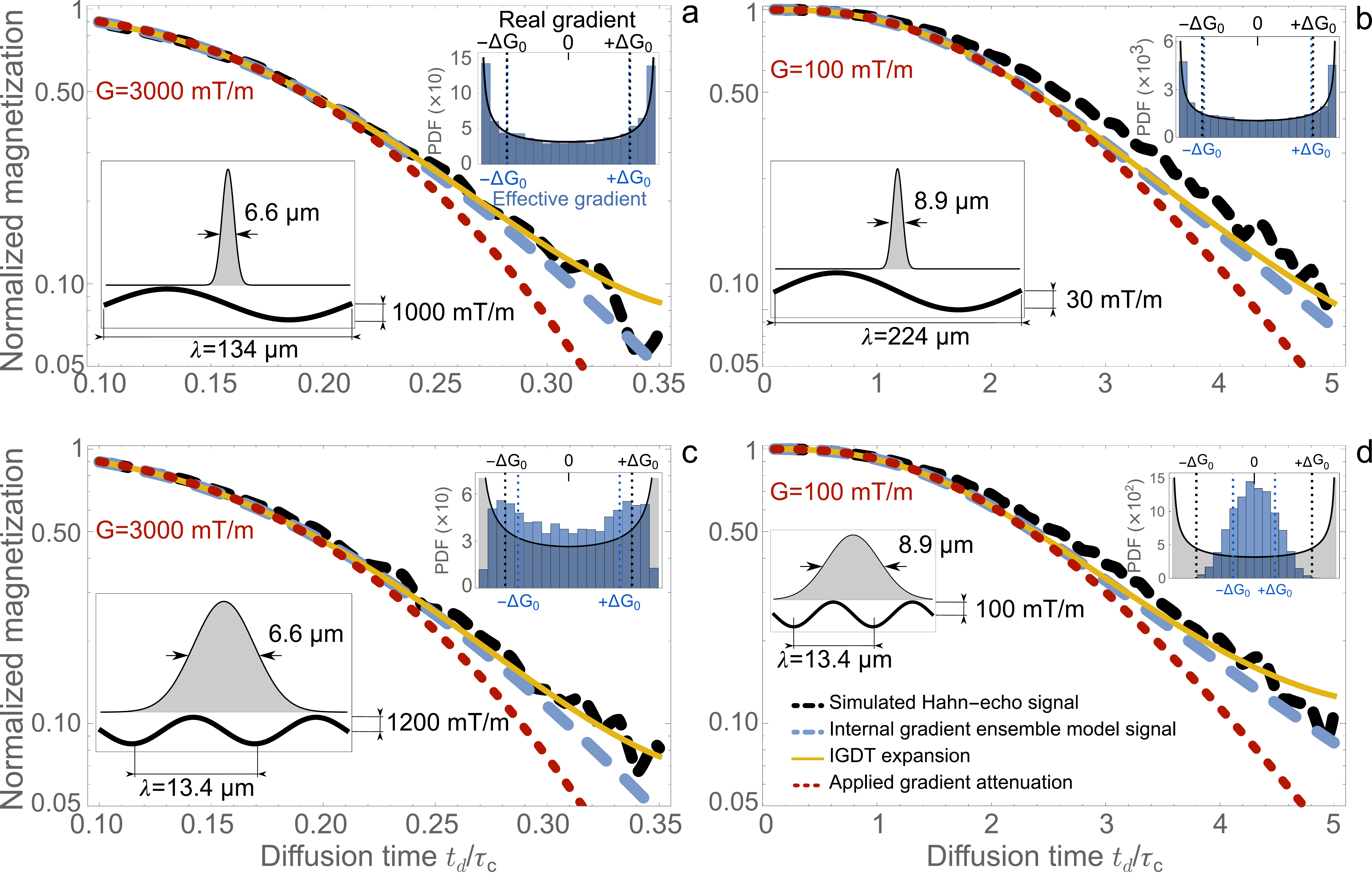}

\caption{\label{fig:simulations}Comparison of the predicted signal derived
from the internal gradient ensemble model and the cumulant expansion
framework with the exact signal of spin diffusing in a inhomogeneous
magnetic field. Simulation of the magnetization decay of spin diffusion
in an inhomogeneous magnetic field shown in the left inset of each
panel in black solid line. The insets also show the relation between
the background internal gradient wavelength and the diffusion length
of the spin-bearing particles determined by the Gaussian distribution
of the spin positions at the maximum diffusion times plotted in the
main panels {[}$t_{d}=3.5$ ms (a,c) and 50 ms (b,d){]}. The insets
on the right of each panel compare the real internal gradient distribution
(black solid curve) with the effective one seen by the spins due to
the motional averaging (blue histogram). Their corresponding standard
deviations are shown with dashed vertical lines. The black dashed
line in the main panels corresponds to the simulated magnetization
for the considered spin-echo sequence of an ensemble of $5\times10^{3}$
spin-bearing particles driven by an OU process in presence of the
sinusoidal background magnetic field gradient of wavelength $\lambda$
and maximum internal gradient strength $g$. The blue dashed line
corresponds to the magnetization averaged with the effective internal
gradient distribution in the right insets using the internal gradient
ensemble model. The red dotted line shows the signal decay only due
to the spin interaction with the applied gradient. The yellow solid
line is the signal decay derived from the IGDT expansion of Eq. (\ref{eq:1D_IGDT_expan}),
evaluated with the moments of the effective gradient distribution
of the right insets. The free diffusion coefficient is $D_{0}=2\,\mathrm{\mu m^{2}/ms}$
and the diffusion correlation time $\tau_{c}=10\,\mathrm{ms}$ (associated
with a restriction length $l_{c}=\sqrt{D_{0}\tau_{c}}=4.47\,\mathrm{\mu m}$).
Four conditions are shown: (a) free diffusion regime with $\lambda=30l_{c}$,
$g=1000\,\mathrm{mT/m}$ and an applied gradient $G=3000\,\mathrm{mT/m}$;
(b) restricted diffusion regime with $\lambda=50l_{c}$, $g=30\,\mathrm{mT/m}$
and $G=100\,\mathrm{mT/m}$; (c) free diffusion regime with $\lambda=3l_{c}$,
$g=1200\,\mathrm{mT/m}$ and $G=3000\,\mathrm{mT/m}$ and (d) restricted
diffusion regime with $\lambda=3l_{c}$, $g=100\,\mathrm{mT/m}$ and
$G=100\,\mathrm{mT/m}$.}
\end{figure*}

The internal gradient distribution that we estimate with the proposed
method is an effective gradient distribution resulting from the motional
averaging of the spin-bearing particles \citep{Huerlimann1998}. The
internal gradient ensemble model implicitly simplifies the signal
attenuation effects due to the spin movement in a space-dependent
gradient field to the one induced by the spin movement in an effective
constant internal gradient. Here, we thus evaluate the reliability
of this assumption through the simulations of the magnetization decay
of spin-bearing particles diffusing in a sinusoidal gradient field
that models typical field inhomogenities in porous systems \citep{Song_2003,Callaghan2011,Fajardo}.

For this evaluation, we perform simulations in one spacial dimension
by setting an applied gradient $G$ and an inhomogeneous background
magnetic field with a $z$-component 
\begin{equation}
\Delta B_{0}(x)=g\frac{\lambda}{2\pi}\cos\left(\frac{2\pi}{\lambda}x\right),\label{eq:back_field}
\end{equation}
where $g$ is the maximum internal gradient strength and $\lambda$
is the wavelength of the inhomogeneous internal field fluctuation.
We consider four scenarios in Fig. \ref{fig:simulations}, free and
restricted diffusion regimes, where the internal magnetic field inhomogenity
wavelength $\lambda$ is larger than and comparable to the diffusion
restriction length $l_{c}$. We do not consider the case when the
wavelength $\lambda$ is shorter than the diffusion restriction length
$l_{c}$, as the signal decay depends only on a averaged internal
gradient value. 

For the simulations, we evaluated 5000 stochastic realizations of
an OU process of mean spin position $x_{0}$ using a free diffusion
coefficient $D_{0}=2\:\mathrm{\mu m^{2}/ms}$ and a diffusion correlation
time $\tau_{c}=10\,\mathrm{ms}$. The initial spin position $x_{0}$
is set randomly within the spatial interval $(-\lambda,\lambda)$.
We then calculate the spin magnetization signal of the diffusing spins
in presence of the considered internal magnetic field inhomogenity,
together with a constant applied gradient with $\pi$-pulse at the
diffusion time $t_{d}/2$ to provide a spin-echo modulation sequence.

The simulations are shown in Fig. \ref{fig:simulations}, where the
black dashed curves are the exact predicted signal. The blue dashed
lines are the magnetization signal predicted by the internal gradient
ensemble model. It is calculated using Eq. (\ref{eq:M_asym}) and
determined by the effective gradient distribution resulting from the
motional averaging of the gradient seen by the spin from the beginning
of the sequence to the total diffusion time 
\begin{equation}
G_{0}^{i}=\frac{1}{t_{d}}\int_{0}^{t_{d}}dt\frac{d}{dx}\Delta B_{0}\left(x_{i}\left(t\right)\right).\label{eq:eff_back_grad}
\end{equation}
Here, the magnetic field shift $\Delta B_{0}\left(x\right)$ is given
by Eq. (\ref{eq:back_field}), $G_{0}^{i}$ is the effective internal
gradient seen by the $i$th spin-bearing particle along the $x_{i}(t)$
position trajectory. With the effective internal gradients $G_{0}^{i}$,
we then generate a histogram as shown in blue in the insets of Fig.
\ref{fig:simulations} compared with the real gradient distribution
in black. The red dotted curves correspond to the magnetization decay
only due to the interaction with the applied gradient and the yellow
solid lines correspond to the predicted signal derived from the IGDT
expansion of Eq. (\ref{eq:1D_IGDT_expan}). In the four considered
cases the expected signals derived from the cumulant expansion and
internal gradient ensemble model are in good agreement with the simulated,
exact one. Thus, this result shows that the effects in the magnetization
decay of spin-bearing particles, diffusing in internal gradients that
depend on its position, can be reduced to an effective internal gradient
distribution based on an ensemble of gradients. Moreover, the magnetization
signal decay is shown to be characterized only by the moments of the
effective internal gradient distribution.

The presented simulations thus show the relevance and good signal
prediction of the developed model to describe the complex diffusion
process in presence of magnetic field inhomogenities induced by magnetic
susceptibility heterogeneity in porous systems. Yet, we envisage some
limitations. While the strength of the model is to reduce the complexity
of the internal magnetic field heterogenitity to an internal gradient
ensemble model characterized by the main moments of the distribution,
it can only determine those moments and not the full internal gradient
distribution. Thus, for example, the method is unable to provide information
about multimodal gradient distributions. The molecular diffusion process
can narrow the effective gradient distribution seen by the spins around
its mean value as shown in Fig. \ref{fig:simulations}d. Then, depending
on the real gradient distribution variance, if the diffusion mean
squared displacement is on the order of $\lambda$, the effective
gradient variance may be much smaller than the real one or even the
IGDT effects may vanish. While the predicted signal might be in agreement
with the exact signal, as in Fig. \ref{fig:simulations}d, some quantitative
information about gradient field heterogeneities might be lost. However,
while this still has to be demostrated, this limitation is probably
a physical limitation due to the motional averaging, and not due to
the assumed model.

Another limitation is imposed by the time domain where the cumulant
expansion is valid, defined by its convergence radius $t_{conv}$
\citep{Kiselev2007,Jones2010}. The cumulant expansion of Eq. (\ref{eq:cum_exp_phi})
is valid for times shorter than its convergence radius, defined by
$t_{conv}=|z|$ with $z$ being the complex solution of $M(z)=0$
and $M(z)$ the analytic continuation of Eq. (\ref{eq:magnetization_def_1})
in the complex domain. Then, if the standard deviation of the internal
gradient distribution is too high (for example $>500\:\mathrm{mT/m}$
in our simulations) the IGDT expansion would be only valid at short
times within the free diffusion regimen as shown in Fig. \ref{fig:simulations}a
and c.

\section{Conclusions\label{sec:Conclusions}}

Diffusion processes in disordered systems are of great interest in
several areas as physics, biology and medicine, however they are very
complex to characterize. In this article, we exploit the diffusion
of spin-bearing particles to probe magnetic field inhomogenities that
contain information about the media microstructure heterogeneities.
We consider internal gradients induced by magnetic susceptibility
heterogeneity that depend on the morphological information of the
media. The spin-bearing particles diffuse within them and probe the
effective gradients seen along their movement. We introduced a model
that simplifies the naturally complex problem to an ensemble of spatially
constant internal gradients that defines a distribution that can be
characterized by its first moments. We demonstrate that this gradient
ensemble approximation is useful as the gradients explored by the
spin-bearing particles remains close to the real distribution in general,
as in expected realistic conditions. We demonstrate that the diffusion
process in an inhomogeneous field produces a motional average of the
gradient seen by the spins, thus narrowing the effective internal
gradient distribution. We also show that the spin magnetization decay
is well characterized by only the moments of the effective internal
gradient distribution.

This internal gradient ensemble model allows a spin phase's cumulant
expansion decomposition for the magnetization signal decay. Based
on this cumulant expansion framework, we derive an IGDT-expansion
that provides the moments of the internal gradient distributions.
The IGDT contains information about the anisotropies of the internal
gradient distributions that intrinsically depends on the susceptibility
heterogeneity of the media and its orientation with respect to the
static magnetic field. Each IGDT term provides different information
about the internal gradient distribution.

The IGDT effects may be weak in some scenarios, thus exploiting the
presented framework, we propose modulated gradient spin echo sequences
to enhance those effects. The enhancement strategy is based on smooth
applied gradient modulations and a suitable timing interplay between
the time modulating symmetries of the applied and internal gradients.
We identify four groups of terms of the IGDT expansion that can be
measured independently by a suitable design of the pulse sequence
using multiple applied gradient directions in a similar vein as DTI
is estimated. We demonstrate the feasibility of the implementation
of this framework to measure IGDT terms from the magnetization signal
decay in free and restricted diffusion regimes using typical brain
tissue conditions. We also evaluate the validity of the model to represent
the real distribution convoluted with the molecular diffusion process,
and show the regimes where the model can be suitably used. We also
discussed some limitations of the model.

We observed that the effective internal gradient distribution reproduces
well the real one when the diffusion is either free or the characteristic
length of the internal gradients spatial variations is small compared
with the diffusion length of the spin-bearing particles. However,
in the restricted diffusion regime, the internal local gradients might
be averaged by the molecular motion, thus lossing information about
their distribution. We thus expect the proposed IGDT method to be
useful to monitor extra-axonal diffusion, since characteristic length
might be wider than in the intra-axonal regions, thus allowing to
extend the free diffusion time to probe smooth internal gradient variations
\citep{Fajardo}.

As brain physiology is regulated by molecules and structures as the
myelin sheath of axons with significant magnetic susceptibility in
comparison with the surrounding medium, the degree of axon myelination
significantly affects the internal gradient distributions \citep{Holland1986,Liu2011,Lee2012,Fajardo}.
Thus internal gradient distributions show correlations with the amount
of myelin in tissues as potential biomarkers for many degenerative
diseases. Our results thus contribute to estimate IGDT in this situations
that may be especially useful for unveiling structures and fiber orientation
based on these susceptibility induced changes \citep{Chen_2013,Alvarez2017,Xu_2017,Fajardo}.
The IGDT are complementary to DTI as they may estimate anisotropies
when diffusion is free or isotropic \citep{Han2011,Alvarez2017,Fajardo}.

\section*{Declaration of interests}

The authors declare that they have no known competing financial interests
or personal relationships that could have appeared to influence the
work reported in this paper.
\begin{acknowledgments}
We thank Melisa Gimenez and Martin Kuffer for reading the article
and for their suggestions to improve the writing, and to Manuel O.
Caceres for the fruitful discussions. This work was supported by CNEA;
CONICET; ANPCyT-FONCyT PICT-2017-3156, PICT-2017-3699, PICT-2018-4333,
PICT-2021-GRF-TI-00134, PICT-2021-I-A-00070; PIP-CONICET (11220170100486CO);
UNCUYO SIIP Tipo I 2019-C028, 2022-C002, 2022-C030; Instituto Balseiro;
A collaboration program from MINCyT (Argentina) and MAECI (Italy)
and Erasmus+ Higher Education program from the European Commission
between the CIMEC (University of Trento) and the Instituto Balseiro
(Universidad Nacional de Cuyo).
\end{acknowledgments}

\appendix

\section{\label{sec:General-demonstration}General demonstration of the proposed
sequence design conditions defines $|\boldsymbol{\beta}^{0G}(t_{d})|-\boldsymbol{\beta}^{GG}(t_{d})$
positive-definite}

We show in Sec. \ref{subsec:Sequence-design} of the main text, a
necessary relation between the overlap matrices $\boldsymbol{\beta}^{GG}$
and $\boldsymbol{\beta}^{0G}$ that have to be fulfilled for enhancing
the cross term contributions in Eq. (\ref{eq:cum_expan}). Thus the
relation we derive is a necessary condition, since there are other
factors that control the intensity of these terms such as the angle
between the applied and mean background gradient or even the applied
gradient strength. Considering the first two terms $\frac{1}{2}G_{i}\beta_{ij}^{GG}(t)G_{j}$
and $G_{i}\beta_{ij}^{0G}(t)\left\langle \boldsymbol{G}_{0}\right\rangle _{j}$
in Eq. (\ref{eq:cum_expan}), we require 
\begin{equation}
\left|\boldsymbol{n}\cdot\boldsymbol{\beta}^{0G}\cdot\boldsymbol{n}\right|>\boldsymbol{n}\cdot\boldsymbol{\beta}^{GG}\cdot\boldsymbol{n}\label{eq:appendix_0GnnGGnn}
\end{equation}
for the matrix elements in an arbitrary $n$ direction, to enhance
the weight of the second term compared to the first one. We thus focus
on the maximization of the cross terms overlap matrices $\boldsymbol{\beta}^{0G}$.
In the principal axes of the matrices $\boldsymbol{\beta}^{GG}$ and
$\boldsymbol{\beta}^{0G}$, the matrix element can be written as 
\[
\left|n\cdot\boldsymbol{\beta}^{0G}\cdot n\right|=\left|\sum_{j}\left(\tilde{n}{}_{j}\right)^{2}\beta_{j}^{0G}\right|\leq\sum_{j}\left(\tilde{n}{}_{j}\right)^{2}\left|\beta_{j}^{0G}\right|,
\]
where $\tilde{n}{}_{j}$ are the components of the unit vector\textbf{
$\boldsymbol{n}$} in the overlap matrices\textbf{ }eigenbasis. Thus
requiring the condition $n\cdot\left|\boldsymbol{\beta}^{0G}\right|\cdot n>n\cdot\boldsymbol{\beta}^{GG}\cdot n$
is equivalent to request $|\boldsymbol{\beta}^{0G}(t_{d})|-\boldsymbol{\beta}^{GG}(t_{d})$
to be positive-definite.

In section \ref{subsec:Sequence-design}, we derived general conditions
that the gradient control sequences have to fulfill in order to satisfy
(\ref{eq:beta_0G > beta_GG}). That is: \emph{i}) the background gradient
modulation $f_{0}(t)$ has to be an even function with respect to
$t_{d}/2$ with a minimal number of pulses; \emph{ii}) the applied
gradient modulation has to be an MGSE sequence with a smooth modulation
function ($|f_{G}(t)|\leq1$), with a single refocusing echo matching
the zero crossing time with a $\pi$-pulse that modulate $f_{0}(t)$
and \emph{iii}) the modulation function $f_{G}(t)$ has to be an odd
function with respect to its middle time, \emph{i.e.} the zero crossing
time, and should vanish also at every $\pi$-pulse that modulate $f_{0}(t)$.
These conditions ensure that the effective interaction times defined
in Eqs. (\ref{eq:TG}) and (\ref{eq:T0G}) satisfy $T_{0G}>T_{G}$.
Moreover, if $\omega_{0}$ is the frequency at the highest pick of
the cross modulation filter and $\omega_{G}$ the frequency at the
maximum applied gradient filter, the above conditions also ensure
$\omega_{G}>\omega_{0}$.

The filters in the overlap integrals in Eqs. (\ref{eq:betaGG}) and
(\ref{eq:beta0G}) have a bandwidth around the frequencies $\omega_{G}$
and $\omega_{0}$, respectively. If the filter bandwidths are small
compared with the spectral density width, we can factor out from the
integral the noise spectral density at the filter frequency, since
it is approximately constant within the filter bandwidth. Then, Eqs.
(\ref{eq:betaGG}) and (\ref{eq:beta0G}) can be approximated by $\boldsymbol{\beta}^{GG}(t_{d})\approx\gamma^{2}\boldsymbol{S}(\omega_{G})T_{G}$
and $\boldsymbol{\beta}^{0G}(t_{d})\approx\gamma^{2}\boldsymbol{S}(\omega_{0})T_{0G}$,
respectively. If not, we can roughly hold this approximation, since
we can take the constant $\boldsymbol{S}(\omega^{*})$ out from the
integral, with $\omega^{*}$ belonging to the filter bandwidth and
$\omega^{*}\approx\omega_{G}$ and $\omega^{*}\approx\omega_{0}$
respectively. Here, the diffusion time $t_{d}$ dependence is encoded
in the effective interactions times and gradient modulation frequencies.
Generally, the eigenvalues $S(\omega)$ of the spectral density are
monotonically decreasing functions of $\omega$ with a maximum at
zero frequency. Then, since $\omega_{G}>\omega_{0}$ and $T_{0G}>T_{G}$,
$|\boldsymbol{\beta}^{0G}(t_{d})|-\boldsymbol{\beta}^{GG}(t_{d})$
is positive-definite.

\section{\label{sec:Optimizing-the-cross-filter's}Enhancing the cross-filter
overlap with the diffusion spectral density}

In section \ref{subsec:Sequence-design}, we derive the conditions
that gradient modulation sequences have to fulfill to guarantee the
matrix $|\boldsymbol{\beta}^{0G}(t_{d})|-\boldsymbol{\beta}^{GG}(t_{d})$
to be positive-definite. This condition is ensured by making the effective
interaction times of gradients with the diffusing spin-bearing particles
defined in Eqs. (\ref{eq:TG}, \ref{eq:T0G}) to satisfy $T_{0G}>T_{G}$.
This condition is achieved by making $f_{0}(t)f_{G}(t)=|f_{G}(t)|$.

We thus aim to maximize the eigenvalues of the matrix $|\boldsymbol{\beta}^{0G}(t_{d})|-\boldsymbol{\beta}^{GG}(t_{d})$.
For that, we maximize the lowest frequency peak of the cross-filter
$\Re\left[F_{0}(\omega,t_{d})F_{G}^{*}(\omega,t_{d})\right]$. In
principle, the condition $f_{0}(t)f_{G}(t)=|f_{G}(t)|$ and the maximization
of the low frequency peak of the cross-filter are independents, and
might not be satisfied simultaneously. Here we demonstrate that, in
fact, both condition could be satisfied at the same time.

First we evaluate the condition for the time-shift of the applied
gradient modulation for the asymmetric sequence that maximize \emph{the
cross-term} overlap contribution. This time-shift maximizes the cross-filter
at the background gradient modulation frequency in order to optimize
the overlap with the displacement spectral density $\boldsymbol{S}(\omega)$.
The cross-filter can be written as 
\begin{equation}
\Re\left[F_{0}(\omega)F_{G}^{*}(\omega)\right]=\widetilde{F}_{0}(\omega)\widetilde{F}_{G}(\omega)\cos\left[\omega\Delta t-\pi/2\right]\label{eq:cross-filter_optimo}
\end{equation}
assuming an arbitrary time-shift $\Delta t$, where $\widetilde{F}_{0}(\omega)$
and $\widetilde{F}_{G}(\omega)$ are real functions corresponding
to the Fourier transform of the gradient modulations centered at $t=0$.
The phase shift in $\pi/2$ comes from the fact that we consider the
applied gradient modulation as an odd function. Here we omit the diffusion
time $t_{d}$ dependence of the Fourier transform in order to simplify
the notation. Let be $\omega_{0}$ the principal harmonic frequency
of the background gradient modulation $f_{0}(t)$. For the two pulse
CPMG sequence, the main harmonic is $\omega_{0}\approx2\pi/t_{d}$.
To optimize the cross-filter of Eq. (\ref{eq:cross-filter_optimo}),
we require that 
\[
\omega_{0}\Delta t-\pi/2=n\pi
\]
 with $|n|=0,1,2...$. Then, the optimal time-shift is
\begin{equation}
\Delta t=\frac{\pi}{\omega_{0}}(n+1/2).\label{eq: time_shift}
\end{equation}
 There are two solution with $n$ such that $\Delta t<t_{d}/2$ which
give $\Delta t=\pm t_{d}/4$. These time shifts synchronize exactly
the zero crossing time of $f_{G}(t)$ with the times when the RF $\pi$-pulse
are applied in the two pulse CPMG sequence.

This optimization is useful as long as $|\widetilde{F}_{0}(\omega_{0})|>|\widetilde{F}_{G}(\omega_{0})|>0$.
The first relation is always true by construction and the second one
impose a restriction on how high can be the frequency with which can
be modulated the applied gradient. Thus we show how both requirements
to enhance the background gradient effect, the effective interaction
times relation $T_{0G}>T_{G}$ and the low frequency filtering of
the cross modulation, can be optimized simultaneously.

\section{\label{sec:Filters-and-overlap}General filters and overlap functions}

In this section we provide the exact expression for the filters of
the GDGSE modulation, a two-pulses CPMG sequence modulation, and the
cross filter $\mathfrak{R}\left[F_{0}(\omega,t_{d})F_{G}^{*}(\omega,t_{d})\right]$
that appear in Eqs. (\ref{eq:betaGG}-\ref{eq:beta0G}) in the main
text. We consider the gradients modulations independent of the applied
gradient direction. We also provide the exact analytic expressions
for the corresponding overlap functions for one dimensional diffusion.
The filter function for the two pulse CPMG sequence of the background
gradient modulation is 
\begin{equation}
|F_{0}(\omega,t_{d})|^{2}=\frac{256}{\omega^{2}}\cos^{2}\left(\frac{t_{d}\omega}{8}\right)\sin^{6}\left(\frac{t_{d}\omega}{8}\right).
\end{equation}
The Fourier transform of the GDGSE modulation is 
\begin{equation}
F_{G}(\omega,t_{d})=i\sqrt{2\pi e}\alpha^{2}t_{d}^{2}\omega e^{i\frac{T}{2}\omega}e^{-\frac{\alpha^{2}t_{d}^{2}}{2}\omega^{2}}\label{eq: FG}
\end{equation}
and thus its filter function is
\begin{equation}
|F_{G}(\omega,t_{d})|^{2}=2\pi e\alpha^{4}t_{d}^{4}\omega^{2}e^{-\alpha^{2}t_{d}^{2}\omega^{2}}.\label{eq: filtro FG}
\end{equation}
 The cross-filter is 
\begin{align}
\mathfrak{R}\left[F_{0}(\omega,t_{d})F_{G}^{*}(\omega,t_{d})\right] & =32\sqrt{2\pi e}\alpha^{2}t_{d}^{2}e^{-\frac{\alpha^{2}t_{d}^{2}}{2}\omega^{2}}\nonumber \\
 & \times\cos^{2}\left(\frac{t_{d}\omega}{8}\right)\sin^{4}\left(\frac{t_{d}\omega}{8}\right).
\end{align}
 With these filters and considering the diffusion spectral density
\[
S(\omega)=2D_{0}\left(\tau_{c}^{-2}+\omega^{2}\right)^{-1},
\]
 we determine the overlap functions using Eqs. (\ref{eq:betaGG}-\ref{eq:beta0G})
in the main text
\begin{align}
\beta^{GG}(t_{d}) & =2\gamma^{2}D_{0}\sqrt{\pi}e\alpha^{3}t_{d}^{3}\nonumber \\
 & \times\left[1-\frac{\sqrt{\pi}\alpha t_{d}}{\tau_{c}}e^{\frac{\alpha^{2}t_{d}^{2}}{\tau_{c}^{2}}}\textrm{erfc}\left(\frac{\alpha t_{d}}{\tau_{c}}\right)\right],\label{eq:beta_GG_full}
\end{align}
 
\begin{align}
\beta^{00}(t_{d}) & =2\gamma^{2}D_{0}\tau_{c}^{3}\left(\frac{t_{d}}{\tau_{c}}-5+e^{-\frac{t_{d}}{\tau_{c}}}-4e^{-\frac{3t_{d}}{4\tau_{c}}}\right.\nonumber \\
 & \left.+4e^{-\frac{t_{d}}{2\tau_{c}}}+4e^{-\frac{t_{d}}{4\tau_{c}}}\right),\label{eq:beta_00_full}
\end{align}
 and
\begin{align}
\beta^{0G}(t_{d})= & \gamma^{2}D_{0}\sqrt{\frac{\pi e}{2}}\tau_{c}\alpha^{2}t_{d}^{2}e^{\frac{\alpha^{2}t_{d}^{2}}{2\tau_{c}^{2}}}\left[4\textrm{erfc}\left(\frac{\alpha t_{d}}{\sqrt{2}\tau_{c}}\right)\right.\nonumber \\
 & -2e^{\frac{t_{d}}{2\tau_{c}}}\textrm{erfc}\left(\frac{\alpha t_{d}/\tau_{c}+1/(2\alpha)}{\sqrt{2}}\right)\nonumber \\
 & -2e^{-\frac{T}{2\tau}}\textrm{erfc}\left(\frac{\alpha t_{d}/\tau_{c}-1/(2\alpha)}{\sqrt{2}}\right)\nonumber \\
 & -e^{\frac{t_{d}}{4\tau_{c}}}\textrm{erfc}\left(\frac{\alpha T/\tau_{c}+1/(4\alpha)}{\sqrt{2}}\right)\nonumber \\
 & -e^{-\frac{t_{d}}{4\tau_{c}}}\textrm{erfc}\left(\frac{\alpha t_{d}/\tau_{c}-1/(4\alpha)}{\sqrt{2}}\right)\nonumber \\
 & +e^{\frac{3t_{d}}{4\tau_{c}}}\textrm{erfc}\left(\frac{\alpha t_{d}/\tau_{c}+3/(4\alpha)}{\sqrt{2}}\right)\nonumber \\
 & \left.+e^{-\frac{3t_{d}}{4\tau_{c}}}\textrm{erfc}\left(\frac{\alpha t_{d}/\tau_{c}-3/(4\alpha)}{\sqrt{2}}\right)\right].\label{eq:beta_0G_full}
\end{align}
 Here, $\textrm{erfc}(x)$ denote the complementary error function.
The generalization to a three-dimensional diffusion is made by considering
the correlation time $\tau_{c}$ as a rank 3 tensor, whose eigenvalues
are the correlation times in its principal directions, as is shown
in Fig \ref{fig:RW3D} of the main text.

\bibliographystyle{elsarticle-num}
\bibliography{IGDT_LAPP_GAA}

\end{document}